\documentclass[aps,prd,twocolumn,a4paper,superscriptaddress]{revtex4-1}
\usepackage{graphicx,multirow,latexsym}
\begin{document}

\newcommand{\be}{\begin{equation}}
\newcommand{\ee}{\end{equation}}
\newcommand{\bq}{\begin{eqnarray}}
\newcommand{\eq}{\end{eqnarray}}
\newcommand{\bsq}{\begin{subequations}}
\newcommand{\esq}{\end{subequations}}
\newcommand{\bc}{\begin{center}}
\newcommand{\ec}{\end{center}}
\newcommand{\al}{\alpha}

\title{Sub-percent constraints on cosmological temperature evolution}

\author{A. Avgoustidis}
\email[]{Anastasios.Avgoustidis@nottingham.ac.uk}
\affiliation{School of Physics and Astronomy, University of Nottingham, University Park, Nottingham NG7 2RD, England}
\author{R. T. G{\'e}nova-Santos}
\email[]{rgs@iac.es}
\affiliation{Instituto de Astrof{\'\i}sica de Canarias, C/V{\'\i}a L{\'a}ctea s/n, La Laguna, Tenerife, Spain}
\affiliation{Departamento de Astrof\'{\i}sica, Universidad de La Laguna (ULL), 38206 La Laguna, Tenerife, Spain}
\author{G. Luzzi}
\email[]{gemma.luzzi@roma1.infn.it}
\affiliation{Dept. of Physics, Sapienza, University of Rome, Piazzale Aldo Moro 2, I-00185 Rome,
Italy}
\author{C. J. A. P. Martins}
\email[]{Carlos.Martins@astro.up.pt}
\affiliation{Centro de Astrof\'{\i}sica, Universidade do Porto, Rua das Estrelas, 4150-762 Porto, Portugal}
\affiliation{Instituto de Astrof\'{\i}sica e Ci\^encias do Espa\c co, CAUP, Rua das Estrelas, 4150-762 Porto, Portugal}

\begin{abstract}
The redshift dependence of the cosmic microwave background temperature is one of the key cosmological observables. In the standard cosmological model one has $T(z)=T_0(1+z)$, where $T_0$ is the present-day temperature. Deviations from this behavior would imply the presence of new physics. Here we discuss how the combination of all currently available direct and indirect measurements of $T(z)$ constrains the common phenomenological parametrization $T(z)=T_0(1+z)^{1-\beta}$, and obtain the first sub-percent constraint on the $\beta$ parameter, specifically $\beta=(7.6\pm8.0)\times10^{-3}$ at the $68.3\%$ confidence level.
\end{abstract}

\pacs{}
\keywords{}
\preprint{}
\maketitle

\section{Introduction}
\label{intro}

Over the last three decades Cosmology has been transformed from a purely theoretical into an observational discipline.  This 
has been possible thanks to a plethora of different observables, notably the Cosmic Microwave Background (CMB) anisotropies 
at $z\!\simeq\!1100$ \cite{cpp2015-13}, the Baryon Acoustic Oscillations (BAO) seen in the distribution of galaxies at 
$z\!\simeq\! 0.5$ \cite{anderson14} or the type Ia supernovae which demonstrated the accelerated expansion of the 
Universe \cite{riess,perlmutter99}. These have led to the consolidation of the so-called $\Lambda$CDM ``concordance model'', 
according to which the Universe is homogeneous on large scales, has a nearly flat geometry, is currently undergoing accelerated 
expansion, and is made of dark energy (68\%), cold dark matter (27\%) and ordinary matter (5\%). 

Despite the extraordinary success of finding completely independent observables converging into a common theoretical framework, there are 
still aspects that are not fully understood or well characterized. Probably the most striking one is that 95\% of the content of the Universe
has not so far been experimentally detected in the laboratory (but has only been detected `mathematically'); this is 
in the form of dark energy and dark matter. This fact strongly hints at the existence of new physics beyond the standard $\Lambda$CDM model. 
In this context, it is important to explore laboratory or astrophysical probes that may provide evidence for the presence of this, still 
unknown, physics. In the present paper we focus on testing the redshift-dependence of the CMB temperature, which is one of the core predictions 
of standard Big Bang cosmology that may be violated under non-standard scenarios \cite{Tofz}. 

According to the Big Bang model, the CMB temperature evolves with redshift $z$ as $T_{\rm CMB}(z)=T_0(1+z)$, under the assumptions of 
adiabatic expansion and photon number conservation. There are however many non-standard scenarios where these assumptions are not met (we will point out
examples of them later) causing potentially observable deviations from the standard scaling. Therefore, direct or indirect 
measurements of the temperature-redshift relation 
provide constraints on scenarios beyond the standard $\Lambda$CDM paradigm. As will be discussed in Section~\ref{direct}, there are several 
ways of obtaining direct constraints on $T(z)$, and these can be combined with indirect constraints coming from measurements of the so-called 
distance duality relation, presented in Section~\ref{indirect}. In the future, further indirect constraints will become available, for example from CMB 
spectral distortions \cite{chluba14}. In Section~\ref{discussion} we present joint constraints after combining the direct and indirect measurements of 
Sections~\ref{direct} and \ref{indirect}. This updates the previous results of \cite{Tofz} and improves them by almost a factor of two,
reaching sub-percent precision for the first time.  Finally, in Section~\ref{conc} we present the main conclusions derived from this study.

\section{Direct constraints from CMB-temperature Measurements}
\label{direct}

Deviations of the standard CMB temperature scaling with redshift are usually described using the parametrization proposed by \cite{lima00},
\be\label{Tofz_beta}
T_{\rm CMB}(z) = T_0 (1+z)^{1-\beta}\,,
\ee 
where $\beta$ is a constant parameter ($\beta=0$ in the standard scenario). The COBE-FIRAS experiment observations provided the 
most-precise blackbody spectrum ever measured, with a temperature at the present epoch, $z=0$, of $T_0=2.7260 \pm 0.0013$~K \cite{fixsen09}. 
At higher redshifts, there are presently two main methods used to obtain direct estimates of $T_{\rm CMB}$, and from which constraints on $\beta$ 
can be derived. The first method we will use was proposed nearly 40 years ago \cite{fabbri78,rephaeli80} and is based on multi-frequency 
observations of the 
Sunyaev-Zel'dovich (SZ) effect \cite{sunyaev70}, a distortion of the CMB spectrum produced towards galaxy clusters. As pointed out by \cite{demartino12}, 
the existing large galaxy cluster catalogues together with very precise CMB data should allow precisions on $\beta$ of the order of $0.01$, a notable 
improvement with respect to initial constrains using a few clusters \cite{battistelli02,luzzi09}. Recent results based on data from the Planck satellite 
\cite{demartino15,luzzi15,hurier14} and from the {\it South Pole Telescope} \cite{saro14} are shown in Table~\ref{tab:constraints_tcmb}. The most precise 
determinations are those from \cite{demartino15} and \cite{luzzi15}, and were obtained using respectively 481 and 103 galaxy clusters. 

\begin{table*}
\begin{center}
\begin{tabular}{ccccccccccccc}
\hline\hline
\noalign{\smallskip}
Method &~& Reference &~& $z$ &~& $N$ &~& $T_{\rm CMB}$ (K) &~~& $\beta$ && Label \\
\noalign{\smallskip}\hline\noalign{\smallskip}
 \multirow{19}{*}{\rotatebox{90}{SZ effect towards clusters}} && \multirow{2}{*}{Saro et al. (2014) \cite{saro14}} && $0.055-1.350$ && 158 &&- &&$0.017\pm  0.030$ && [a]\\     
  && && $0.3-1.350$ &&  &&- &&  $0.016\pm 0.031$  && [b]\\
\noalign{\smallskip}\cline{3-13}\noalign{\smallskip}
             && de Martino et al. (2015) \cite{demartino15} && $<0.3$  && 481  && - && $-0.007\pm 0.013$ && [c]\\
\noalign{\smallskip}\cline{3-13}\noalign{\smallskip}
              && \multirow{3}{*}{Luzzi et al. (2015) \cite{luzzi15}} && $0.011-0.972$ && 103 && - && $0.012\pm 0.016$ && [d]\\
              &&                                               && $0.011-0.972$     && 99 && - && $0.014\pm 0.016$  && [e]\\
              &&                                               && $0.3-0.972$     && 33 && - && $0.020\pm 0.017$  && [f]\\      
\noalign{\smallskip}\cline{3-13}\noalign{\smallskip}
              && \multirow{3}{*}{Luzzi et al. (2009) \cite{luzzi09}} && $0.023-0.546$ && 13 && - && $0.065\pm 0.080 $ && [g]\\
              &&                                               && $0.200-0.546$     && 7 && - && $0.044\pm 0.087$  && [h]\\	 
              &&                                               && $0.3-0.546$     && 2 && - && $0.05\pm 0.14$  && [i]\\      
\noalign{\smallskip}\cline{3-13}\noalign{\smallskip}
&& \multirow{13}{*}{Hurier et al. (2014) \cite{hurier14}} && $0-1$  && 813  &&  - &&  $0.009\pm 0.017$ && [j] \\
\noalign{\smallskip}\cline{5-13}\noalign{\smallskip}
&& && $0.30 -0.35$  && 81  &&  $3.562\pm 0.050$ &&  \multirow{12}{*}{$-0.006\pm 0.022$} && \multirow{12}{*}{[k]} \\
&& && $0.35 -0.40$  && 50  &&  $3.717\pm 0.063$ &&  && \\
&& && $0.40 -0.45$  && 45  &&  $3.971\pm 0.071$ &&   &&\\
&& && $0.45 -0.50$  && 26  &&  $3.943\pm 0.112$ &&   &&\\
&& && $0.50 -0.55$  && 20  &&  $4.380\pm 0.119$ &&   &&\\
&& && $0.55 -0.60$  && 18  &&  $4.075\pm 0.156$ &&   &&\\
&& && $0.60 -0.65$  && 12  &&  $4.404\pm 0.194$ &&   &&\\
&& && $0.65 -0.70$  &&  6  &&  $4.779\pm 0.278$ &&   &&\\
&& && $0.70 -0.75$  &&  5  &&  $4.933\pm 0.371$ &&   &&\\
&& && $0.75 -0.80$  &&  2  &&  $4.515\pm 0.621$ &&   &&\\
&& && $0.85 -0.90$  &&  1  &&  $5.356\pm 0.617$ &&   &&\\
&& && $0.95 -1.00$  &&  1  &&  $5.813\pm 1.025$ &&   &&\\
\noalign{\smallskip}\hline\noalign{\smallskip}
 \multirow{15}{*}{\rotatebox{90}{QSO absorption lines}}  &&  Muller et al. (2013) \cite{muller13} &&  $0.89$  && 1 && $5.0791_{-0.0994}^{+0.0993}$ &&  \multirow{15}{*}{$0.005\pm 0.022$}&& \multirow{15}{*}{[l]}\\  
 \noalign{\smallskip}\cline{3-9}\noalign{\smallskip}
&& \multirow{3}{*}{Noterdeame et al. (2011) \cite{noterdaeme11}} && $1.7293$ && 1 &&  $7.5_{-1.2}^{+1.6}$   && &&\\
&& && $1.7738$ && 1 &&  $7.8_{-0.6}^{+0.7}$   && &&\\
&& && $2.0377$ && 1 &&  $8.6_{-1.0}^{+1.1}$   && &&\\
\noalign{\smallskip}\cline{3-9}\noalign{\smallskip}
&&  Cui et al. (2005) \cite{cui05}&&  $1.77654$  && 1 && $7.2\pm 0.8$ && &&\\ 
\noalign{\smallskip}\cline{3-9}\noalign{\smallskip}
&&  Ge et al. (2001) \cite{ge01} &&  $1.9731$  && 1 && $7.9\pm 1.0$ &&  &&\\ 
\noalign{\smallskip}\cline{3-9}\noalign{\smallskip}
&&  Srianand et al. (2000) \cite{srianand00} &&  $2.33771$  && 1 && $6-14$ &&  &&\\ 
\noalign{\smallskip}\cline{3-9}\noalign{\smallskip}
&& Srianand (2008) \cite{srianand08} && $2.4184$ && 1 &&  $9.15\pm 0.72$  && &&\\
\noalign{\smallskip}\cline{3-9}\noalign{\smallskip}
&& Noterdaeme et al. (2010) \cite{noterdaeme10} && $2.6896$ && 1 &&  $10.5_{-0.6}^{+0.8}$  && &&\\
\noalign{\smallskip}\cline{3-9}\noalign{\smallskip}
&&  Molaro et al. (2002) \cite{molaro02} &&  $3.025$  && 1 && $12.1_{-3.2}^{+1.7}$ &&  &&\\ 
\noalign{\smallskip}
\hline\hline
\end{tabular}
\caption{\label{tab:constraints_tcmb} Measurements of the CMB taken from the literature, derived from the SZ effect towards galaxy 
clusters, and from CMB-photon induced rotational excitation of CO, C{\sc i} and C{\sc ii} in quasar spectral lines. 
$N$ is the number of objects that were used and is different 
from unity only in cases of combining SZ observations towards many galaxy clusters. In those cases we indicate the range of redshifts of the clusters. 
The fifth column shows the derived CMB temperature, and the last column the derived constraints on the $\beta$ parameter describing the 
CMB-temperature redshift evolution. For all the measurements derived from SZ studies we show the $\beta$ values which have been taken directly from the corresponding references and also the $\beta$ values we have recalculated after removing the overlapping clusters among the various samples. In particular, in the case of Luzzi et al. (2015), we quote the final $\beta$ constraint using their full sample of $N=103$ clusters, and also our recalculation after removing the four clusters ($N=99$) in common with Saro et al. (2014), and after keeping all clusters with $z>0.3$ and removing the  
only cluster in common with Saro et al. (2014) in that redshift range ($N=33$).
In the case of Luzzi et al. (2009), we quote the final $\beta$ constraint using their sample of $N=13$ clusters, and also our recalculation 
after removing the six clusters ($N=7$) in common with Luzzi et al. (2015) [e], and after keeping only clusters with $z>0.3$ ($N=2$).
While Hurier et al. (2014) give a $\beta$ constraint using their full cluster sample between $z=0$ and $z=1$, here we estimate $\beta$ using only their 
$T_{\rm CMB}(z)$ values between $z=0.3$ and $z=1$, in such a way that this constraint can be complemented with the one from de Martino et al. 
(2015). The same reanalysis has been applied to the Saro et al. (2014) sample. 
We have also estimated the $\beta$ constraint using several $T_{\rm CMB}(z)$ measurements in quasar spectral 
lines.}
\end{center}
\normalsize
\medskip
\end{table*}

In order to combine different measurements we have to ensure that they are independent, and therefore we have to remove any overlapping clusters. 
The analysis of \cite{demartino15} uses an unpublished catalogue containing clusters at $z<0.3$, so to combine with their measurement we will remove all clusters 
in this redshift range from the other samples. In Table~\ref{tab:constraints_tcmb} we show the $\beta$ values resulting from the \citet{luzzi15} subsamples 
containing: 
\begin{itemize} 

\item  99 clusters after removing the 4 clusters that are in 
common with the SPT sample in their full redshift range, and 
\item 33 clusters at $z>0.3$ and removing one cluster in common with the SPT sample. 
\end{itemize}
We have the original posterior distributions of $T_{\rm CMB}(z)$ for each of the clusters of \cite{luzzi15}, 
which have been used in this reanalysis. Similarly \cite{hurier14} obtained $\beta=0.009\pm 0.017$ using 813 clusters out to $z\approx 1$.
In Table~\ref{tab:constraints_tcmb} we show the constraint we have derived using only their clusters at $z>0.3$. To this end we have used their 
$T_{\rm CMB}(z)$ values obtained after stacking clusters in $\Delta z=0.05$ redshift bins, and assuming Gaussian distributions. We also show 
the result of our reanalysis of the \citet{saro14} sample using only their $z>0.3$ clusters. 

Estimations of $T_{\rm CMB}(z)$ through the SZ effect are currently limited to $z\lesssim 1$ due to the scarcity of galaxy clusters at high redshifts. Estimates 
at $z>1$ can be obtained through the study of quasar absorption line spectra which show energy levels that have been excited through atomic or 
molecular transitions after the absorption of CMB photons \cite{bahcall68}. If the relative populations of the different energy levels are in radiative 
equilibrium with the CMB radiation, then the excitation temperature gives the temperature of the CMB at that redshift. Early estimates based 
on this method must be regarded as upper limits on $T_{\rm CMB}(z)$, since there could be significant contributions from other local sources of 
excitation.  The first constraints using this method were only obtained 15 years ago \cite{srianand00}, taking advantage of the enormous progress in high-resolution astrophysical
spectroscopy; they use transitions in the UV range due to the excitation of fine-structure levels 
of atomic species like C{\sc i} or C{\sc ii} \cite{srianand00,ge01,molaro02,cui05}. More recently, improved constraints have been obtained 
from precise measurements of CO transitions and radio-mm transitions produced by the rotational excitation of molecules
with permanent dipole moment  \cite{srianand08,noterdaeme10,noterdaeme11,muller13}. In Table~\ref{tab:constraints_tcmb} we show all these $T_{\rm CMB}(z)$ estimates 
and our derived joint constraint $\beta=0.005\pm 0.022$.

It must be noted that recently the Planck Collaboration~\cite{cpp2015-13} obtained a very stringent constraint of $\beta=0.0004\pm 0.0011$ by combining CMB with large-scale 
structure data, after fixing the recombination redshift at $z_\star = 1100$. The very low error bar on $\beta$ is due to the long lever-arm 
in redshift afforded by the CMB. However, that constraint only applies to models were the deviation from adiabatic evolution starts at the last scattering 		
surface and, perhaps more importantly, the parametrization they use, while adequate for low redshifts is completely unrealistic for $z\sim1000$. 
We leave the discussion of physically motivated high-redshift parametrizations for subsequent work.

Here we derive stringent constraints by combining the SZ and QSO absorption measurements shown in Table~\ref{tab:constraints_tcmb}. 
These two techniques complement each other not only because they cover different redshift ranges but also because they are subject to different 
types of systematics. Despite the shortage of targets, spectroscopic observations cover redshifts out to $z\approx 3$, therefore providing a longer lever-arm, as 
opposed to SZ observations that are restricted to $z<1$ but benefit from a larger number of targets provided by the SZ cluster catalogues recently 
published \cite{cpp2015-27}.

\begin{figure*}[t]
\includegraphics[width=16cm]{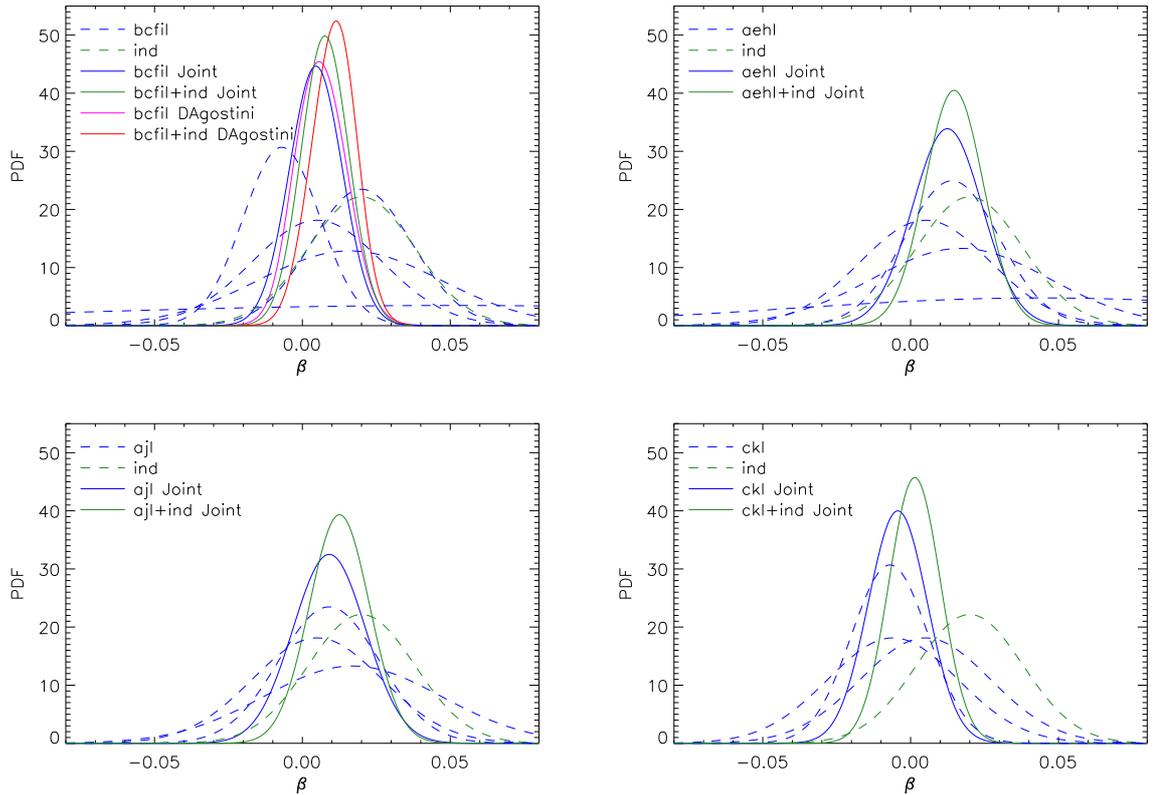}
\caption{\label{figpdf} Individual and joint probability density functions for each of the four combinations indicated in Tables~\ref{tab:joint} and \protect\ref{tab:joint_indirect}. Blue dashed lines show the individual PDFs derived from the different direct constraints on the $\beta$ parameter shown in Table~\ref{tab:constraints_tcmb}, assuming Gaussian error distributions. The green dashed lines correspond to the indirect constraint derived in section~\protect\ref{indirect}. The solid blue lines show the joint constraints from direct measurements, from which the expected values and confidence intervals of Table~\ref{tab:joint} have been derived. The green solid lines correspond to the combination of direct and indirect constraints, and are associated to the values of Table~\protect\ref{tab:joint_indirect}. Finally, the magenta and red lines represent also a joint constraint, but obtained using the prescription of \protect\cite{Dagos1999}.} 
\end{figure*}

In Table~\ref{tab:joint} we present different possible combinations avoiding overlapping clusters, and the resulting joint constraints on $\beta$, which have been obtained by a standard weighted mean combination. In Figure~\ref{figpdf}, the blue dashed lines represent the probability density functions (PDFs), assumed to be Gaussian, corresponding to different combinations presented in Table~\ref{tab:constraints_tcmb}. The solid blue lines represent the joint PDFs, which are just the multiplication of the individual PDFs. Due to a marginal disagreement among the data combined in case [b]+[c]+[f]+[i]+[l], we have also applied the formula for a `skeptical' combination of experimental results, proposed by D'Agostini~\cite{Dagos1999}. This PDF is represented by the magenta line in the top-left panel of Figure~\ref{figpdf}. As soon as the individual results start to disagree, the combined distribution gets broader with respect to the standard (weighted mean) result. However, if the agreement among individual results is good the combined distribution becomes narrower than the standard result.
In this case we find that the expected value is slightly shifted from the one obtained by the simple weighted mean. The intent of using this formula is to take into account the dispersion of the measurements, nevertheless this method returns lower errors. This is due to the fact that there are 4 of 5 measurements which are in mutual agreement and one in marginal disagreement. The application of this formula results in a higher weight to the data sample in agreement, thus causing the shift of the expected value but not the broadening of the distribution.

\begin{table}
\begin{center}
\begin{tabular}{ccc}
\hline\hline
\noalign{\smallskip}
Combination &~& $\beta$\\
\noalign{\smallskip}\hline\noalign{\smallskip}
[b]+[c]+[f]+[i] +[l]&& $0.0046\pm 0.0089$ \footnote{Applying the prescription of \protect\cite{Dagos1999} we get $0.0064\pm 0.0086$}\\
\noalign{\smallskip}
[a]+[e]+[h]+[l] && $0.012\pm 0.012 $\\
\noalign{\smallskip}
[a]+[j]+[l] && $0.009\pm 0.012$\\
\noalign{\smallskip}
[c]+[k]+[l] && $-0.004\pm 0.010$\\
\noalign{\smallskip}
\hline\hline
\end{tabular}
\caption{\label{tab:joint}Joint constraints on the $\beta$ parameter, derived through the combination of different constraints shown in 
Table~\protect\ref{tab:constraints_tcmb}. The first column indicates the specific combination, represented by the labels listed 
in the last column of Table~\protect\ref{tab:constraints_tcmb}. Individual and joint probably density functions are also plotted in Fig. \protect\ref{figpdf}.}
\end{center}
\normalsize
\medskip
\end{table}

\section{Indirect constraints from Distance Measurements}  
\label{indirect}

As discussed in \cite{Tofz}, indirect constraints on the evolution $T(z)$, which are complementary (and competitive) to 
the direct bounds discussed above, can be obtained from the comparison between different distance measurements at 
the same redshift. In the standard cosmological picture, photon number is conserved and so luminosity distances should 
agree with other distance measures, as for example angular diameter or radial $H(z)$ 
distance determinations. However, if photon number conservation is fundamentally -- or effectively -- violated, there will 
be a systematic mismatch between luminosity distances that depend crucially on conservation of photon number, and 
other distance measures that are not sensitive to photon number conservation. 

Any deviation from the standard picture 
in which photons can decay or be absorbed or emitted along the line of sight would give rise to such a breakdown 
of ``distance duality" \cite{BassettKunz}. Examples include couplings of photons with axions and axion-like scalar 
fields \cite{Csaki_PRL,Csaki_PLB}, or other particles beyond the standard model (see \cite{ABRVJ} and references 
therein), phenomenological models of Dark Energy interacting with photons \cite{Bur_Cham,AMMVL}, a hypothetical 
grey dust \cite{Aguirre}, or intergalactic dust \cite{Menard}. Note that in the case of couplings between axion-like scalars and 
photons the overall effect can also lead to an apparent brightening of the source, for example if axions are also emitted at the 
source and subsequently decay to photons along the line of sight.         

The potential mismatch between different distance determinations due to any of the above effects can be readily constrained 
at the percent level with current data. Traditionally this has been done by constraining possible violations of the so-called 
Etherington (or distance duality) relation \cite{Etherington} through the parametrization:
\be\label{dLdA_eps}
d_L(z) = d_A(z) (1+z)^{2+\epsilon}\,.
\ee  
Here, $\epsilon$ simply parametrizes deviations from the standard relation between luminosity distance $d_L$ and angular 
diameter distance $d_A$. This parametrization is not physically motivated but it is adequate at low redshifts $z \lesssim 1$
for which $\epsilon z \ll 1$, thus matching the first term in a Taylor expansion, proportional to $\epsilon z$. As more 
data at larger redshifts are now becoming available, a more appropriate parametrization is required. When particular physical 
models are considered, for example SN dimming/brightening due to couplings of photons with axions or other particles beyond  
the standard model, specific parametrizations can be used \cite{ABRVJ}, guided by the physics of each model. However, generic 
bounds are usually quoted in terms of the parameter $\epsilon$.      
   
If such a violation of the distance duality relation was due to fundamental interactions between optical photons from Supernovae 
and some other field permeating space, one would expect that the same field would also interact with CMB photons, causing for 
example spectral distortions \cite{MirRafSerp_CMB} (see also \cite{Ellis2013,Brax2013}). In reference \cite{Tofz} the authors pointed out 
that, since these couplings could also cause deviations from the standard temperature evolution law discussed above (\ref{Tofz_beta}), 
constraints in the parameters $\epsilon$ in (\ref{dLdA_eps}) and $\beta$ in (\ref{Tofz_beta}) should be explicitly related within 
a given model. In particular, for the simplest possible case of adiabatic achromatic dimming of the CMB, the relation between $\beta$ and $\epsilon$ is:
\be\label{betaepsrel}
\beta = -\frac{2}{3} \epsilon \,.
\ee
Further, in \cite{AMMVL} these potential deviations from the standard $T(z)$ law and the distance duality relation were linked to variations 
of constants of nature, in particular the fine structure constant $\alpha$. Assuming that the $\alpha$ variation is due to a linear gauge kinetic
function (a well motivated scenario, as discussed in \cite{Dvali}) and further using adiabatic achromatic dimming as a toy example, one 
finds a simple linear relation between $\epsilon$, $\beta$ and $\Delta\alpha/\alpha$. Such relations are of course model-dependent 
and can be more complicated in realistic models. In \cite{Hees14} the authors did this calculation for generic non-minimal multiplicative
couplings between a scalar field and the matter sector, which produce 
non-adiabatic dimming. This confirms the simple linear relation between $\beta$ and $\Delta\alpha/\alpha$ to lowest order, correcting 
the relevant coefficient by a factor of order unity: the coefficient $-2/3$ in Eq. (\ref{betaepsrel}) would change to $-0.24$. Therefore, using
the adiabatic approximation actually yields a more conservative indirect bound on $\beta$ from distance duality tests.    

These parametric relations among violations of standard physical laws (that can be probed with different observables) provide an 
important tool for bootstrapping observational constraints on the cosmological paradigm. Thus spectroscopic and SZ determinations of $T(z)$, SN 
luminosity distances, BAO and $H(z)$ data from galaxy ageing can be optimally combined to cross-check constraints and break
degeneracies. They provide an exciting opportunity to probe fundamental High-Energy Physics interactions like scalar-photon couplings
suppressed by energy scales of order $\sim 10^{10}$ GeV, using low energy observations of $\sim\!$ eV scale photons in the 
late universe.  

We first update previous indirect constraints on $\beta$ using the latest available distance determinations. 
We use the SDSS-II/SNLS3 JLA sample~\cite{Betoule2014} for luminosity distances and compare 
with a number of different determinations of $H(z)$: cosmic chronometers \cite{Stern2009,Simon2004,Jimenez2003} 
(11 data points in the redshift range $0.1<z<1.75$) and the more recent \cite{Moresco2012} (8 data points at 
$0.17<z<1.1$), BAO combined with Alcock-Paczynski (AP) distortions to separate the radial component 
in the WiggleZ Dark Energy Survey \cite{Blake2012} (3 data points at $z=0.44,0.6$ and $0.73$), the 
SDSS DR7 BAO measurement \cite{Xu2012} at $z=0.35$, the BOSS DR11 measurement 
\cite{Anderson2013} at $z=0.57$, and the recent $H(z)$ determination at z=2.3 from BAO in the Ly$\alpha$ 
forest of BOSS DR11 quasars \cite{Delubac2014}. This gives 25 data points in the range $0.1<z<2.3$.  
 
\begin{figure}
\includegraphics[height=2.7in,width=3.0in]{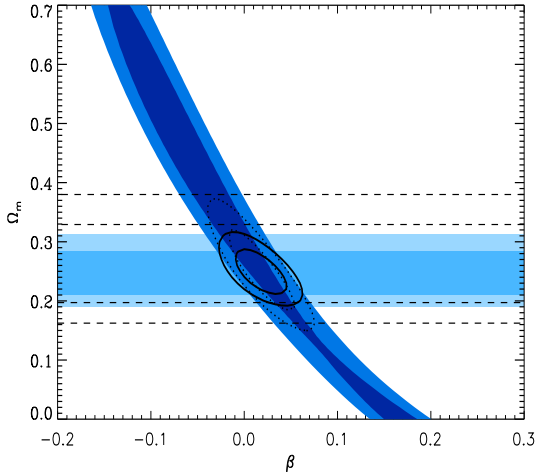}
  \caption{\label{beta_constr} Constraints from SN+H(z) on the parameter $\beta$, parameterising 
  violations of the temperature-redshift relation as $T(z) = T_0(1 + z)^{1-\beta}$. Dark blue contours 
  correspond to 68\% and 95\% confidence levels obtained from SN data alone, light blue contours 
  are for H(z) data, and solid line transparent contours show the joint SN+H(z) constraint. There is 
  significant improvement from the previous constraint of reference \protect\cite{Tofz} (dotted contours), 
  which comes mainly from the inclusion of more $H(z)$/BAO datapoints compared to the ``cosmic 
  chronometers'' determinations \protect\cite{Stern2009} (dashed lines) used in \protect\cite{Tofz}.} 
\end{figure} 
    
Figure~\ref{beta_constr} shows our joint constraint on the $\beta-\Omega_m$ plane. We have taken 
flat $\Lambda$CDM models, marginalised over $H_0$, and have assumed the simple relation (\ref{betaepsrel}) 
between $\beta$ and $\epsilon$ in (\ref{Tofz_beta}) and (\ref{dLdA_eps}) respectively. Marginalising over $\Omega_m$ gives:
\be\label{beta_ind}
\beta = 0.020 \pm 0.018 \ \ \ (1\sigma)
\ee
which is competitive to the most stringent constraints from the direct probes of Section~\ref{direct} and is subject 
to completely different systematics. As we will see in the next section, the combination of this constraint with direct 
bounds on $\beta$ yields the first sub-percent constraints on the cosmic temperature-redshift evolution.

\section{Data Combination and Discussion}
\label{discussion}

The combination of direct and indirect bounds on the parameter $\beta$ discussed in Sections~\ref{direct} 
and~\ref{indirect} above leads to a significant improvement of the overall constraint on $\beta$ with an uncertainty
reaching sub-percent levels. In Table~\ref{tab:joint_indirect} we show the joint constraints corresponding to 
Table~\ref{tab:joint} but now also including the indirect constraint coming from distance measurements.  The PDF 
of this indirect constraint is represented by the green dashed lines in Figure~\ref{figpdf}, while the final joint PDFs 
resulting form the combination of direct and indirect measurements are depicted by the green solid lines.

\begin{table}
\begin{center}
\begin{tabular}{ccc}
\hline\hline
\noalign{\smallskip}
Combination &~& $\beta$\\
\noalign{\smallskip}\hline\noalign{\smallskip}
[b]+[c]+[f]+[l]+[i]+[indirect]  && $0.0076\pm 0.0080$ \footnote{Applying the prescription of \protect\cite{Dagos1999} we get $0.0106\pm 0.0076$}\\
\noalign{\smallskip}
[a]+[e]+[l]+[h]+[indirect]  && $0.0147\pm 0.0099$\\
\noalign{\smallskip}
[a]+[j]+[l]+[indirect]  && $0.013\pm 0.010$\\
\noalign{\smallskip}
[c]+[k]+[l]+[indirect]  && $0.0014\pm 0.0087$ \\
\noalign{\smallskip}
\hline\hline
\end{tabular}
\caption{\label{tab:joint_indirect}Joint constraints on the $\beta$ parameter, derived through the combination of different constraints shown in Table~\protect\ref{tab:constraints_tcmb} and the indirect constraints of Section~\protect\ref{indirect}. The first column indicates the specific combination, represented by the labels listed in the last column of Table~\protect\ref{tab:constraints_tcmb}. The label [indirect] corresponds to the constraint (\ref{beta_ind}) of Section~\ref{indirect}. Individual and joint probably density functions are also plotted in Fig. \protect\ref{figpdf}.}
\end{center}
\normalsize
\medskip
\end{table}

Two comments are in order regarding the interpretation and model-dependence of these results. Direct determinations of $T(z)$ 
are subject to systematic uncertainties which have been included in the errors we have used in our analysis. On the other hand, the 
link between distance measurement constraints and bounds on deviations from the temperature-redshift relation is model-dependent. 
In particular, when connecting constraints coming from Supernovae observations (optical photons) to deviations of the CMB temperature 
from its standard form (probed at much longer wavelengths), one implicitly assumes that the dimming mechanism is wavelength-independent. 
This is a strong assumption, but a plausible one over a wide range of photon frequencies, for example in the context of 
axion-photon couplings. Further, on general grounds, couplings of this type are expected to be weaker for lower photon frequencies 
so assuming a frequency-independent coupling as we did yields conservative bounds on $T(z)$ violations from SN data. 

The parametrizations we have used in (\ref{Tofz_beta}) and (\ref{dLdA_eps}) to quantify deviations from the standard 
temperature-redshift and luminosity-angular diameter distance relationships are standard in the literature, so constraints are better 
expressed in terms of these parameters to facilitate direct comparisons with other bounds in the literature (refer to Sections \ref{direct} 
and \ref{indirect}). However, they are phenomenological and are not directly derived from any particular theoretical model. Since deviations 
from the standard relations are constrained to be small, and are currently mostly probed down to redshifts of order unity, these
parametrizations are adequate at present. In particular, Taylor-expanding equations (\ref{Tofz_beta}) and (\ref{dLdA_eps}) in redshift allows 
comparison to physical variables in any given model. As more data at larger redshifts are gradually becoming available, the lowest order Taylor 
approximation breaks down (at $z>$ few) and more accurate parametrizations are needed. This can be done in a model-to-model basis. For example, 
in the case of SN dimming/brightening due to couplings of photons with axions or other particles beyond the standard model, specific parametrizations 
can be used \cite{ABRVJ}, which are guided by the physics of each model. The simple parametrizations used in this work remain useful as a 
phenomenological way to study deviations from the standard $T(z)$ and distance duality relationships, without referring to a specific physical 
model.

\section{Conclusions}   
\label{conc}

We have revisited existing constraints on deviations from the adiabatic evolution of the CMB black-body temperature, and by combining the latest direct (thermal SZ effect and precision spectroscopy) and indirect (distance measures) probes, we obtained the first sub-percent constraint on such deviations, parametrized by the simple phenomenological law given by Eq.~(\ref{Tofz_beta}). Namely, we have found $\beta=(7.6\pm8.0)\times10^{-3}$ at the $68.3\%$ confidence level (Table~\ref{tab:joint_indirect}). These measurements provide an important consistency test of the standard cosmological model (as any deviation from adiabaticity will imply the presence of new physics) and also provide an important external dataset for other cosmological probes such as Euclid \cite{AMMVL}.

We note that although Eq.~(\ref{Tofz_beta}) is a reasonable parametrization at low redshifts (specifically, for $z\lesssim 1$), it is not expected to be realistic for larger redshift ranges, in the sense that physically motivated models will typically lead to a different behavior in the matter era. We have used it in the present work for the simple reason that it is the canonical one in the currently available literature, and therefore it allows the results of our analysis to be easily compared with those of earlier works. Nevertheless it is already clear that as the quality, quantity and redshift span of the data improves more realistic classes of models should be tested.

Very significant improvements are expected in the coming years. The next generation of space-based CMB missions (e.g., a COrE/PRISM-like mission \cite{PRISM}) may improve the number of available SZ measurements of the CMB temperature by as much as two orders of magnitude, although detailed simulations of the impact of such a dataset remain to be done. As for spectroscopic measurements, ALMA and ESPRESSO will soon be making significant contributions \cite{ALMA1,ALMA2,ESPRESSO} and the prospects are even better for the high resolution ultra-stable spectrograph at the European Extremely Large Telescope \cite{EELT}. In this case most of the progress is expected to come from CO measurements which are signal-to-noise limited. (CN would provide an even better thermometer, but so far this has not been detected in high-redshift absorption systems \cite{KREL}.) A roadmap for these measurements and a discussion of their role in precision consistency tests of the standard model can be found in \cite{GRG}.

\begin{acknowledgments}
This work was supported by Funda\c c\~ao para a Ci\^encia e a Tecnologia (FCT) through the research grants PTDC/FIS/111725/2009 and UID/FIS/04434/2013. CJM is also supported by an FCT Research Professorship, contract reference IF/00064/2012, funded by FCT/MCTES (Portugal) and POPH/FSE (EC). AA is supported by the University of Nottingham Research Board through an NRF Fellowship. This work was partially supported by funding from University of Rome Sapienza 2014 C26A14FP3T. We thank A. Saro for sharing the data points plotted on figure 1 of~\cite{saro14} from the SPT analysis. 
\end{acknowledgments}

\bibliography{Tofz1pc}

\begin{thebibliography}{59}%
\makeatletter
\providecommand \@ifxundefined [1]{%
 \@ifx{#1\undefined}
}%
\providecommand \@ifnum [1]{%
 \ifnum #1\expandafter \@firstoftwo
 \else \expandafter \@secondoftwo
 \fi
}%
\providecommand \@ifx [1]{%
 \ifx #1\expandafter \@firstoftwo
 \else \expandafter \@secondoftwo
 \fi
}%
\providecommand \natexlab [1]{#1}%
\providecommand \enquote  [1]{``#1''}%
\providecommand \bibnamefont  [1]{#1}%
\providecommand \bibfnamefont [1]{#1}%
\providecommand \citenamefont [1]{#1}%
\providecommand \href@noop [0]{\@secondoftwo}%
\providecommand \href [0]{\begingroup \@sanitize@url \@href}%
\providecommand \@href[1]{\@@startlink{#1}\@@href}%
\providecommand \@@href[1]{\endgroup#1\@@endlink}%
\providecommand \@sanitize@url [0]{\catcode `\\12\catcode `\$12\catcode
  `\&12\catcode `\#12\catcode `\^12\catcode `\_12\catcode `\%12\relax}%
\providecommand \@@startlink[1]{}%
\providecommand \@@endlink[0]{}%
\providecommand \url  [0]{\begingroup\@sanitize@url \@url }%
\providecommand \@url [1]{\endgroup\@href {#1}{\urlprefix }}%
\providecommand \urlprefix  [0]{URL }%
\providecommand \Eprint [0]{\href }%
\providecommand \doibase [0]{http://dx.doi.org/}%
\providecommand \selectlanguage [0]{\@gobble}%
\providecommand \bibinfo  [0]{\@secondoftwo}%
\providecommand \bibfield  [0]{\@secondoftwo}%
\providecommand \translation [1]{[#1]}%
\providecommand \BibitemOpen [0]{}%
\providecommand \bibitemStop [0]{}%
\providecommand \bibitemNoStop [0]{.\EOS\space}%
\providecommand \EOS [0]{\spacefactor3000\relax}%
\providecommand \BibitemShut  [1]{\csname bibitem#1\endcsname}%
\let\auto@bib@innerbib\@empty
\bibitem [{\citenamefont {{Planck Collaboration}}\ \emph
  {et~al.}(2015{\natexlab{a}})\citenamefont {{Planck Collaboration}},
  \citenamefont {{Ade}}, \citenamefont {{Aghanim}}, \citenamefont {{Arnaud}},
  \citenamefont {{Ashdown}}, \citenamefont {{Aumont}}, \citenamefont
  {{Baccigalupi}}, \citenamefont {{Banday}}, \citenamefont {{Barreiro}},
  \citenamefont {{Bartlett}},\ and\ \citenamefont {et~al.}}]{cpp2015-13}%
  \BibitemOpen
  \bibfield  {author} {\bibinfo {author} {\bibnamefont {{Planck
  Collaboration}}}, \bibinfo {author} {\bibfnamefont {P.~A.~R.}\ \bibnamefont
  {{Ade}}}, \bibinfo {author} {\bibfnamefont {N.}~\bibnamefont {{Aghanim}}},
  \bibinfo {author} {\bibfnamefont {M.}~\bibnamefont {{Arnaud}}}, \bibinfo
  {author} {\bibfnamefont {M.}~\bibnamefont {{Ashdown}}}, \bibinfo {author}
  {\bibfnamefont {J.}~\bibnamefont {{Aumont}}}, \bibinfo {author}
  {\bibfnamefont {C.}~\bibnamefont {{Baccigalupi}}}, \bibinfo {author}
  {\bibfnamefont {A.~J.}\ \bibnamefont {{Banday}}}, \bibinfo {author}
  {\bibfnamefont {R.~B.}\ \bibnamefont {{Barreiro}}}, \bibinfo {author}
  {\bibfnamefont {J.~G.}\ \bibnamefont {{Bartlett}}}, \ and\ \bibinfo {author}
  {\bibnamefont {et~al.}},\ }\href@noop {} {\bibfield  {journal} {\bibinfo
  {journal} {ArXiv e-prints}\ } (\bibinfo {year} {2015}{\natexlab{a}})},\
  \Eprint {http://arxiv.org/abs/1502.01589} {arXiv:1502.01589} \BibitemShut
  {NoStop}%
\bibitem [{\citenamefont {{Anderson}}\ \emph {et~al.}(2014)\citenamefont
  {{Anderson}}, \citenamefont {{Aubourg}}, \citenamefont {{Bailey}},
  \citenamefont {{Beutler}}, \citenamefont {{Bhardwaj}}, \citenamefont
  {{Blanton}}, \citenamefont {{Bolton}}, \citenamefont {{Brinkmann}},
  \citenamefont {{Brownstein}}, \citenamefont {{Burden}}, \citenamefont
  {{Chuang}}, \citenamefont {{Cuesta}}, \citenamefont {{Dawson}}, \citenamefont
  {{Eisenstein}}, \citenamefont {{Escoffier}}, \citenamefont {{Gunn}},
  \citenamefont {{Guo}}, \citenamefont {{Ho}}, \citenamefont {{Honscheid}},
  \citenamefont {{Howlett}}, \citenamefont {{Kirkby}}, \citenamefont
  {{Lupton}}, \citenamefont {{Manera}}, \citenamefont {{Maraston}},
  \citenamefont {{McBride}}, \citenamefont {{Mena}}, \citenamefont
  {{Montesano}}, \citenamefont {{Nichol}}, \citenamefont {{Nuza}},
  \citenamefont {{Olmstead}}, \citenamefont {{Padmanabhan}}, \citenamefont
  {{Palanque-Delabrouille}}, \citenamefont {{Parejko}}, \citenamefont
  {{Percival}}, \citenamefont {{Petitjean}}, \citenamefont {{Prada}},
  \citenamefont {{Price-Whelan}}, \citenamefont {{Reid}}, \citenamefont
  {{Roe}}, \citenamefont {{Ross}}, \citenamefont {{Ross}}, \citenamefont
  {{Sabiu}}, \citenamefont {{Saito}}, \citenamefont {{Samushia}}, \citenamefont
  {{S{\'a}nchez}}, \citenamefont {{Schlegel}}, \citenamefont {{Schneider}},
  \citenamefont {{Scoccola}}, \citenamefont {{Seo}}, \citenamefont {{Skibba}},
  \citenamefont {{Strauss}}, \citenamefont {{Swanson}}, \citenamefont
  {{Thomas}}, \citenamefont {{Tinker}}, \citenamefont {{Tojeiro}},
  \citenamefont {{Maga{\~n}a}}, \citenamefont {{Verde}}, \citenamefont
  {{Wake}}, \citenamefont {{Weaver}}, \citenamefont {{Weinberg}}, \citenamefont
  {{White}}, \citenamefont {{Xu}}, \citenamefont {{Y{\`e}che}}, \citenamefont
  {{Zehavi}},\ and\ \citenamefont {{Zhao}}}]{anderson14}%
  \BibitemOpen
  \bibfield  {author} {\bibinfo {author} {\bibfnamefont {L.}~\bibnamefont
  {{Anderson}}}, \bibinfo {author} {\bibfnamefont {{\'E}.}~\bibnamefont
  {{Aubourg}}}, \bibinfo {author} {\bibfnamefont {S.}~\bibnamefont {{Bailey}}},
  \bibinfo {author} {\bibfnamefont {F.}~\bibnamefont {{Beutler}}}, \bibinfo
  {author} {\bibfnamefont {V.}~\bibnamefont {{Bhardwaj}}}, \bibinfo {author}
  {\bibfnamefont {M.}~\bibnamefont {{Blanton}}}, \bibinfo {author}
  {\bibfnamefont {A.~S.}\ \bibnamefont {{Bolton}}}, \bibinfo {author}
  {\bibfnamefont {J.}~\bibnamefont {{Brinkmann}}}, \bibinfo {author}
  {\bibfnamefont {J.~R.}\ \bibnamefont {{Brownstein}}}, \bibinfo {author}
  {\bibfnamefont {A.}~\bibnamefont {{Burden}}}, \bibinfo {author}
  {\bibfnamefont {C.-H.}\ \bibnamefont {{Chuang}}}, \bibinfo {author}
  {\bibfnamefont {A.~J.}\ \bibnamefont {{Cuesta}}}, \bibinfo {author}
  {\bibfnamefont {K.~S.}\ \bibnamefont {{Dawson}}}, \bibinfo {author}
  {\bibfnamefont {D.~J.}\ \bibnamefont {{Eisenstein}}}, \bibinfo {author}
  {\bibfnamefont {S.}~\bibnamefont {{Escoffier}}}, \bibinfo {author}
  {\bibfnamefont {J.~E.}\ \bibnamefont {{Gunn}}}, \bibinfo {author}
  {\bibfnamefont {H.}~\bibnamefont {{Guo}}}, \bibinfo {author} {\bibfnamefont
  {S.}~\bibnamefont {{Ho}}}, \bibinfo {author} {\bibfnamefont {K.}~\bibnamefont
  {{Honscheid}}}, \bibinfo {author} {\bibfnamefont {C.}~\bibnamefont
  {{Howlett}}}, \bibinfo {author} {\bibfnamefont {D.}~\bibnamefont {{Kirkby}}},
  \bibinfo {author} {\bibfnamefont {R.~H.}\ \bibnamefont {{Lupton}}}, \bibinfo
  {author} {\bibfnamefont {M.}~\bibnamefont {{Manera}}}, \bibinfo {author}
  {\bibfnamefont {C.}~\bibnamefont {{Maraston}}}, \bibinfo {author}
  {\bibfnamefont {C.~K.}\ \bibnamefont {{McBride}}}, \bibinfo {author}
  {\bibfnamefont {O.}~\bibnamefont {{Mena}}}, \bibinfo {author} {\bibfnamefont
  {F.}~\bibnamefont {{Montesano}}}, \bibinfo {author} {\bibfnamefont {R.~C.}\
  \bibnamefont {{Nichol}}}, \bibinfo {author} {\bibfnamefont {S.~E.}\
  \bibnamefont {{Nuza}}}, \bibinfo {author} {\bibfnamefont {M.~D.}\
  \bibnamefont {{Olmstead}}}, \bibinfo {author} {\bibfnamefont
  {N.}~\bibnamefont {{Padmanabhan}}}, \bibinfo {author} {\bibfnamefont
  {N.}~\bibnamefont {{Palanque-Delabrouille}}}, \bibinfo {author}
  {\bibfnamefont {J.}~\bibnamefont {{Parejko}}}, \bibinfo {author}
  {\bibfnamefont {W.~J.}\ \bibnamefont {{Percival}}}, \bibinfo {author}
  {\bibfnamefont {P.}~\bibnamefont {{Petitjean}}}, \bibinfo {author}
  {\bibfnamefont {F.}~\bibnamefont {{Prada}}}, \bibinfo {author} {\bibfnamefont
  {A.~M.}\ \bibnamefont {{Price-Whelan}}}, \bibinfo {author} {\bibfnamefont
  {B.}~\bibnamefont {{Reid}}}, \bibinfo {author} {\bibfnamefont {N.~A.}\
  \bibnamefont {{Roe}}}, \bibinfo {author} {\bibfnamefont {A.~J.}\ \bibnamefont
  {{Ross}}}, \bibinfo {author} {\bibfnamefont {N.~P.}\ \bibnamefont {{Ross}}},
  \bibinfo {author} {\bibfnamefont {C.~G.}\ \bibnamefont {{Sabiu}}}, \bibinfo
  {author} {\bibfnamefont {S.}~\bibnamefont {{Saito}}}, \bibinfo {author}
  {\bibfnamefont {L.}~\bibnamefont {{Samushia}}}, \bibinfo {author}
  {\bibfnamefont {A.~G.}\ \bibnamefont {{S{\'a}nchez}}}, \bibinfo {author}
  {\bibfnamefont {D.~J.}\ \bibnamefont {{Schlegel}}}, \bibinfo {author}
  {\bibfnamefont {D.~P.}\ \bibnamefont {{Schneider}}}, \bibinfo {author}
  {\bibfnamefont {C.~G.}\ \bibnamefont {{Scoccola}}}, \bibinfo {author}
  {\bibfnamefont {H.-J.}\ \bibnamefont {{Seo}}}, \bibinfo {author}
  {\bibfnamefont {R.~A.}\ \bibnamefont {{Skibba}}}, \bibinfo {author}
  {\bibfnamefont {M.~A.}\ \bibnamefont {{Strauss}}}, \bibinfo {author}
  {\bibfnamefont {M.~E.~C.}\ \bibnamefont {{Swanson}}}, \bibinfo {author}
  {\bibfnamefont {D.}~\bibnamefont {{Thomas}}}, \bibinfo {author}
  {\bibfnamefont {J.~L.}\ \bibnamefont {{Tinker}}}, \bibinfo {author}
  {\bibfnamefont {R.}~\bibnamefont {{Tojeiro}}}, \bibinfo {author}
  {\bibfnamefont {M.~V.}\ \bibnamefont {{Maga{\~n}a}}}, \bibinfo {author}
  {\bibfnamefont {L.}~\bibnamefont {{Verde}}}, \bibinfo {author} {\bibfnamefont
  {D.~A.}\ \bibnamefont {{Wake}}}, \bibinfo {author} {\bibfnamefont {B.~A.}\
  \bibnamefont {{Weaver}}}, \bibinfo {author} {\bibfnamefont {D.~H.}\
  \bibnamefont {{Weinberg}}}, \bibinfo {author} {\bibfnamefont
  {M.}~\bibnamefont {{White}}}, \bibinfo {author} {\bibfnamefont
  {X.}~\bibnamefont {{Xu}}}, \bibinfo {author} {\bibfnamefont {C.}~\bibnamefont
  {{Y{\`e}che}}}, \bibinfo {author} {\bibfnamefont {I.}~\bibnamefont
  {{Zehavi}}}, \ and\ \bibinfo {author} {\bibfnamefont {G.-B.}\ \bibnamefont
  {{Zhao}}},\ }\href {\doibase 10.1093/mnras/stu523} {\bibfield  {journal}
  {\bibinfo  {journal} {M.N.R.A.S.}\ }\textbf {\bibinfo {volume} {441}},\
  \bibinfo {pages} {24} (\bibinfo {year} {2014})},\ \Eprint
  {http://arxiv.org/abs/1312.4877} {arXiv:1312.4877} \BibitemShut {NoStop}%
\bibitem [{\citenamefont {Riess}\ \emph {et~al.}(1998)\citenamefont {Riess}
  \emph {et~al.}}]{riess}%
  \BibitemOpen
  \bibfield  {author} {\bibinfo {author} {\bibfnamefont {A.~G.}\ \bibnamefont
  {Riess}} \emph {et~al.} (\bibinfo {collaboration} {Supernova Search Team}),\
  }\href {\doibase 10.1086/300499} {\bibfield  {journal} {\bibinfo  {journal}
  {Astron.J.}\ }\textbf {\bibinfo {volume} {116}},\ \bibinfo {pages} {1009}
  (\bibinfo {year} {1998})},\ \Eprint {http://arxiv.org/abs/astro-ph/9805201}
  {arXiv:astro-ph/9805201 [astro-ph]} \BibitemShut {NoStop}%
\bibitem [{\citenamefont {{Perlmutter}}\ \emph {et~al.}(1999)\citenamefont
  {{Perlmutter}}, \citenamefont {{Aldering}}, \citenamefont {{Goldhaber}},
  \citenamefont {{Knop}}, \citenamefont {{Nugent}}, \citenamefont {{Castro}},
  \citenamefont {{Deustua}}, \citenamefont {{Fabbro}}, \citenamefont
  {{Goobar}}, \citenamefont {{Groom}}, \citenamefont {{Hook}}, \citenamefont
  {{Kim}}, \citenamefont {{Kim}}, \citenamefont {{Lee}}, \citenamefont
  {{Nunes}}, \citenamefont {{Pain}}, \citenamefont {{Pennypacker}},
  \citenamefont {{Quimby}}, \citenamefont {{Lidman}}, \citenamefont {{Ellis}},
  \citenamefont {{Irwin}}, \citenamefont {{McMahon}}, \citenamefont
  {{Ruiz-Lapuente}}, \citenamefont {{Walton}}, \citenamefont {{Schaefer}},
  \citenamefont {{Boyle}}, \citenamefont {{Filippenko}}, \citenamefont
  {{Matheson}}, \citenamefont {{Fruchter}}, \citenamefont {{Panagia}},
  \citenamefont {{Newberg}}, \citenamefont {{Couch}},\ and\ \citenamefont
  {{Project}}}]{perlmutter99}%
  \BibitemOpen
  \bibfield  {author} {\bibinfo {author} {\bibfnamefont {S.}~\bibnamefont
  {{Perlmutter}}}, \bibinfo {author} {\bibfnamefont {G.}~\bibnamefont
  {{Aldering}}}, \bibinfo {author} {\bibfnamefont {G.}~\bibnamefont
  {{Goldhaber}}}, \bibinfo {author} {\bibfnamefont {R.~A.}\ \bibnamefont
  {{Knop}}}, \bibinfo {author} {\bibfnamefont {P.}~\bibnamefont {{Nugent}}},
  \bibinfo {author} {\bibfnamefont {P.~G.}\ \bibnamefont {{Castro}}}, \bibinfo
  {author} {\bibfnamefont {S.}~\bibnamefont {{Deustua}}}, \bibinfo {author}
  {\bibfnamefont {S.}~\bibnamefont {{Fabbro}}}, \bibinfo {author}
  {\bibfnamefont {A.}~\bibnamefont {{Goobar}}}, \bibinfo {author}
  {\bibfnamefont {D.~E.}\ \bibnamefont {{Groom}}}, \bibinfo {author}
  {\bibfnamefont {I.~M.}\ \bibnamefont {{Hook}}}, \bibinfo {author}
  {\bibfnamefont {A.~G.}\ \bibnamefont {{Kim}}}, \bibinfo {author}
  {\bibfnamefont {M.~Y.}\ \bibnamefont {{Kim}}}, \bibinfo {author}
  {\bibfnamefont {J.~C.}\ \bibnamefont {{Lee}}}, \bibinfo {author}
  {\bibfnamefont {N.~J.}\ \bibnamefont {{Nunes}}}, \bibinfo {author}
  {\bibfnamefont {R.}~\bibnamefont {{Pain}}}, \bibinfo {author} {\bibfnamefont
  {C.~R.}\ \bibnamefont {{Pennypacker}}}, \bibinfo {author} {\bibfnamefont
  {R.}~\bibnamefont {{Quimby}}}, \bibinfo {author} {\bibfnamefont
  {C.}~\bibnamefont {{Lidman}}}, \bibinfo {author} {\bibfnamefont {R.~S.}\
  \bibnamefont {{Ellis}}}, \bibinfo {author} {\bibfnamefont {M.}~\bibnamefont
  {{Irwin}}}, \bibinfo {author} {\bibfnamefont {R.~G.}\ \bibnamefont
  {{McMahon}}}, \bibinfo {author} {\bibfnamefont {P.}~\bibnamefont
  {{Ruiz-Lapuente}}}, \bibinfo {author} {\bibfnamefont {N.}~\bibnamefont
  {{Walton}}}, \bibinfo {author} {\bibfnamefont {B.}~\bibnamefont
  {{Schaefer}}}, \bibinfo {author} {\bibfnamefont {B.~J.}\ \bibnamefont
  {{Boyle}}}, \bibinfo {author} {\bibfnamefont {A.~V.}\ \bibnamefont
  {{Filippenko}}}, \bibinfo {author} {\bibfnamefont {T.}~\bibnamefont
  {{Matheson}}}, \bibinfo {author} {\bibfnamefont {A.~S.}\ \bibnamefont
  {{Fruchter}}}, \bibinfo {author} {\bibfnamefont {N.}~\bibnamefont
  {{Panagia}}}, \bibinfo {author} {\bibfnamefont {H.~J.~M.}\ \bibnamefont
  {{Newberg}}}, \bibinfo {author} {\bibfnamefont {W.~J.}\ \bibnamefont
  {{Couch}}}, \ and\ \bibinfo {author} {\bibfnamefont {T.~S.~C.}\ \bibnamefont
  {{Project}}},\ }\href {\doibase 10.1086/307221} {\bibfield  {journal}
  {\bibinfo  {journal} {Ap. J.}\ }\textbf {\bibinfo {volume} {517}},\ \bibinfo
  {pages} {565} (\bibinfo {year} {1999})},\ \Eprint
  {http://arxiv.org/abs/astro-ph/9812133} {astro-ph/9812133} \BibitemShut
  {NoStop}%
\bibitem [{\citenamefont {Avgoustidis}\ \emph {et~al.}(2012)\citenamefont
  {Avgoustidis}, \citenamefont {Luzzi}, \citenamefont {Martins},\ and\
  \citenamefont {Monteiro}}]{Tofz}%
  \BibitemOpen
  \bibfield  {author} {\bibinfo {author} {\bibfnamefont {A.}~\bibnamefont
  {Avgoustidis}}, \bibinfo {author} {\bibfnamefont {G.}~\bibnamefont {Luzzi}},
  \bibinfo {author} {\bibfnamefont {C.~J. A.~P.}\ \bibnamefont {Martins}}, \
  and\ \bibinfo {author} {\bibfnamefont {A.~M. R. V.~L.}\ \bibnamefont
  {Monteiro}},\ }\href {\doibase 10.1088/1475-7516/2012/02/013} {\bibfield
  {journal} {\bibinfo  {journal} {JCAP}\ }\textbf {\bibinfo {volume} {1202}},\
  \bibinfo {pages} {013} (\bibinfo {year} {2012})},\ \Eprint
  {http://arxiv.org/abs/1112.1862} {arXiv:1112.1862 [astro-ph.CO]} \BibitemShut
  {NoStop}%
\bibitem [{\citenamefont {{Chluba}}(2014)}]{chluba14}%
  \BibitemOpen
  \bibfield  {author} {\bibinfo {author} {\bibfnamefont {J.}~\bibnamefont
  {{Chluba}}},\ }\href {\doibase 10.1093/mnras/stu1260} {\bibfield  {journal}
  {\bibinfo  {journal} {M.N.R.A.S.}\ }\textbf {\bibinfo {volume} {443}},\
  \bibinfo {pages} {1881} (\bibinfo {year} {2014})},\ \Eprint
  {http://arxiv.org/abs/1405.1277} {arXiv:1405.1277} \BibitemShut {NoStop}%
\bibitem [{\citenamefont {{Lima}}\ \emph {et~al.}(2000)\citenamefont {{Lima}},
  \citenamefont {{Silva}},\ and\ \citenamefont {{Viegas}}}]{lima00}%
  \BibitemOpen
  \bibfield  {author} {\bibinfo {author} {\bibfnamefont {J.~A.~S.}\
  \bibnamefont {{Lima}}}, \bibinfo {author} {\bibfnamefont {A.~I.}\
  \bibnamefont {{Silva}}}, \ and\ \bibinfo {author} {\bibfnamefont {S.~M.}\
  \bibnamefont {{Viegas}}},\ }\href {\doibase 10.1046/j.1365-8711.2000.03172.x}
  {\bibfield  {journal} {\bibinfo  {journal} {M.N.R.A.S.}\ }\textbf {\bibinfo
  {volume} {312}},\ \bibinfo {pages} {747} (\bibinfo {year}
  {2000})}\BibitemShut {NoStop}%
\bibitem [{\citenamefont {{Fixsen}}(2009)}]{fixsen09}%
  \BibitemOpen
  \bibfield  {author} {\bibinfo {author} {\bibfnamefont {D.~J.}\ \bibnamefont
  {{Fixsen}}},\ }\href {\doibase 10.1088/0004-637X/707/2/916} {\bibfield
  {journal} {\bibinfo  {journal} {Ap. J.}\ }\textbf {\bibinfo {volume} {707}},\
  \bibinfo {pages} {916} (\bibinfo {year} {2009})},\ \Eprint
  {http://arxiv.org/abs/0911.1955} {arXiv:0911.1955} \BibitemShut {NoStop}%
\bibitem [{\citenamefont {{Fabbri}}\ \emph {et~al.}(1978)\citenamefont
  {{Fabbri}}, \citenamefont {{Melchiorri}},\ and\ \citenamefont
  {{Natale}}}]{fabbri78}%
  \BibitemOpen
  \bibfield  {author} {\bibinfo {author} {\bibfnamefont {R.}~\bibnamefont
  {{Fabbri}}}, \bibinfo {author} {\bibfnamefont {F.}~\bibnamefont
  {{Melchiorri}}}, \ and\ \bibinfo {author} {\bibfnamefont {V.}~\bibnamefont
  {{Natale}}},\ }\href {\doibase 10.1007/BF00651052} {\bibfield  {journal}
  {\bibinfo  {journal} {Astrophys. \& Sp. Sci.}\ }\textbf {\bibinfo {volume}
  {59}},\ \bibinfo {pages} {223} (\bibinfo {year} {1978})}\BibitemShut
  {NoStop}%
\bibitem [{\citenamefont {{Rephaeli}}(1980)}]{rephaeli80}%
  \BibitemOpen
  \bibfield  {author} {\bibinfo {author} {\bibfnamefont {Y.}~\bibnamefont
  {{Rephaeli}}},\ }\href {\doibase 10.1086/158398} {\bibfield  {journal}
  {\bibinfo  {journal} {Ap. J.}\ }\textbf {\bibinfo {volume} {241}},\ \bibinfo
  {pages} {858} (\bibinfo {year} {1980})}\BibitemShut {NoStop}%
\bibitem [{\citenamefont {{Sunyaev}}\ and\ \citenamefont
  {{Zeldovich}}(1970)}]{sunyaev70}%
  \BibitemOpen
  \bibfield  {author} {\bibinfo {author} {\bibfnamefont {R.~A.}\ \bibnamefont
  {{Sunyaev}}}\ and\ \bibinfo {author} {\bibfnamefont {Y.~B.}\ \bibnamefont
  {{Zeldovich}}},\ }\href {\doibase 10.1007/BF00653471} {\bibfield  {journal}
  {\bibinfo  {journal} {Astrophys. \& Sp. Sci.}\ }\textbf {\bibinfo {volume}
  {7}},\ \bibinfo {pages} {3} (\bibinfo {year} {1970})}\BibitemShut {NoStop}%
\bibitem [{\citenamefont {{de Martino}}\ \emph {et~al.}(2012)\citenamefont {{de
  Martino}}, \citenamefont {{Atrio-Barandela}}, \citenamefont {{da Silva}},
  \citenamefont {{Ebeling}}, \citenamefont {{Kashlinsky}}, \citenamefont
  {{Kocevski}},\ and\ \citenamefont {{Martins}}}]{demartino12}%
  \BibitemOpen
  \bibfield  {author} {\bibinfo {author} {\bibfnamefont {I.}~\bibnamefont {{de
  Martino}}}, \bibinfo {author} {\bibfnamefont {F.}~\bibnamefont
  {{Atrio-Barandela}}}, \bibinfo {author} {\bibfnamefont {A.}~\bibnamefont {{da
  Silva}}}, \bibinfo {author} {\bibfnamefont {H.}~\bibnamefont {{Ebeling}}},
  \bibinfo {author} {\bibfnamefont {A.}~\bibnamefont {{Kashlinsky}}}, \bibinfo
  {author} {\bibfnamefont {D.}~\bibnamefont {{Kocevski}}}, \ and\ \bibinfo
  {author} {\bibfnamefont {C.~J.~A.~P.}\ \bibnamefont {{Martins}}},\ }\href
  {\doibase 10.1088/0004-637X/757/2/144} {\bibfield  {journal} {\bibinfo
  {journal} {AP. J.}\ }\textbf {\bibinfo {volume} {757}},\ \bibinfo {eid} {144}
  (\bibinfo {year} {2012})},\ \Eprint {http://arxiv.org/abs/1203.1825}
  {arXiv:1203.1825} \BibitemShut {NoStop}%
\bibitem [{\citenamefont {{Battistelli}}\ \emph {et~al.}(2002)\citenamefont
  {{Battistelli}}, \citenamefont {{De Petris}}, \citenamefont {{Lamagna}},
  \citenamefont {{Melchiorri}}, \citenamefont {{Palladino}}, \citenamefont
  {{Savini}}, \citenamefont {{Cooray}}, \citenamefont {{Melchiorri}},
  \citenamefont {{Rephaeli}},\ and\ \citenamefont {{Shimon}}}]{battistelli02}%
  \BibitemOpen
  \bibfield  {author} {\bibinfo {author} {\bibfnamefont {E.~S.}\ \bibnamefont
  {{Battistelli}}}, \bibinfo {author} {\bibfnamefont {M.}~\bibnamefont {{De
  Petris}}}, \bibinfo {author} {\bibfnamefont {L.}~\bibnamefont {{Lamagna}}},
  \bibinfo {author} {\bibfnamefont {F.}~\bibnamefont {{Melchiorri}}}, \bibinfo
  {author} {\bibfnamefont {E.}~\bibnamefont {{Palladino}}}, \bibinfo {author}
  {\bibfnamefont {G.}~\bibnamefont {{Savini}}}, \bibinfo {author}
  {\bibfnamefont {A.}~\bibnamefont {{Cooray}}}, \bibinfo {author}
  {\bibfnamefont {A.}~\bibnamefont {{Melchiorri}}}, \bibinfo {author}
  {\bibfnamefont {Y.}~\bibnamefont {{Rephaeli}}}, \ and\ \bibinfo {author}
  {\bibfnamefont {M.}~\bibnamefont {{Shimon}}},\ }\href {\doibase
  10.1086/345589} {\bibfield  {journal} {\bibinfo  {journal} {Ap. J. Lett.}\
  }\textbf {\bibinfo {volume} {580}},\ \bibinfo {pages} {L101} (\bibinfo {year}
  {2002})},\ \Eprint {http://arxiv.org/abs/astro-ph/0208027} {astro-ph/0208027}
  \BibitemShut {NoStop}%
\bibitem [{\citenamefont {{Luzzi}}\ \emph {et~al.}(2009)\citenamefont
  {{Luzzi}}, \citenamefont {{Shimon}}, \citenamefont {{Lamagna}}, \citenamefont
  {{Rephaeli}}, \citenamefont {{De Petris}}, \citenamefont {{Conte}},
  \citenamefont {{De Gregori}},\ and\ \citenamefont {{Battistelli}}}]{luzzi09}%
  \BibitemOpen
  \bibfield  {author} {\bibinfo {author} {\bibfnamefont {G.}~\bibnamefont
  {{Luzzi}}}, \bibinfo {author} {\bibfnamefont {M.}~\bibnamefont {{Shimon}}},
  \bibinfo {author} {\bibfnamefont {L.}~\bibnamefont {{Lamagna}}}, \bibinfo
  {author} {\bibfnamefont {Y.}~\bibnamefont {{Rephaeli}}}, \bibinfo {author}
  {\bibfnamefont {M.}~\bibnamefont {{De Petris}}}, \bibinfo {author}
  {\bibfnamefont {A.}~\bibnamefont {{Conte}}}, \bibinfo {author} {\bibfnamefont
  {S.}~\bibnamefont {{De Gregori}}}, \ and\ \bibinfo {author} {\bibfnamefont
  {E.~S.}\ \bibnamefont {{Battistelli}}},\ }\href {\doibase
  10.1088/0004-637X/705/2/1122} {\bibfield  {journal} {\bibinfo  {journal} {Ap.
  J.}\ }\textbf {\bibinfo {volume} {705}},\ \bibinfo {pages} {1122} (\bibinfo
  {year} {2009})},\ \Eprint {http://arxiv.org/abs/0909.2815} {arXiv:0909.2815}
  \BibitemShut {NoStop}%
\bibitem [{\citenamefont {{de Martino}}\ \emph {et~al.}(2015)\citenamefont {{de
  Martino}}, \citenamefont {{G{\'e}nova-Santos}}, \citenamefont
  {{Atrio-Barandela}}, \citenamefont {{Ebeling}}, \citenamefont {{Kashlinsky}},
  \citenamefont {{Kocevski}},\ and\ \citenamefont {{Martins}}}]{demartino15}%
  \BibitemOpen
  \bibfield  {author} {\bibinfo {author} {\bibfnamefont {I.}~\bibnamefont {{de
  Martino}}}, \bibinfo {author} {\bibfnamefont {R.}~\bibnamefont
  {{G{\'e}nova-Santos}}}, \bibinfo {author} {\bibfnamefont {F.}~\bibnamefont
  {{Atrio-Barandela}}}, \bibinfo {author} {\bibfnamefont {H.}~\bibnamefont
  {{Ebeling}}}, \bibinfo {author} {\bibfnamefont {A.}~\bibnamefont
  {{Kashlinsky}}}, \bibinfo {author} {\bibfnamefont {D.}~\bibnamefont
  {{Kocevski}}}, \ and\ \bibinfo {author} {\bibfnamefont {C.~J.~A.~P.}\
  \bibnamefont {{Martins}}},\ }\href {\doibase 10.1088/0004-637X/808/2/128}
  {\bibfield  {journal} {\bibinfo  {journal} {Ap. J.}\ }\textbf {\bibinfo
  {volume} {808}},\ \bibinfo {eid} {128} (\bibinfo {year} {2015})},\ \Eprint
  {http://arxiv.org/abs/1502.06707} {arXiv:1502.06707} \BibitemShut {NoStop}%
\bibitem [{\citenamefont {Luzzi}\ \emph {et~al.}(2015)\citenamefont {Luzzi},
  \citenamefont {G\'enova-Santos}, \citenamefont {Martins}, \citenamefont
  {De~Petris},\ and\ \citenamefont {Lamagna}}]{luzzi15}%
  \BibitemOpen
  \bibfield  {author} {\bibinfo {author} {\bibfnamefont {G.}~\bibnamefont
  {Luzzi}}, \bibinfo {author} {\bibfnamefont {R.~T.}\ \bibnamefont
  {G\'enova-Santos}}, \bibinfo {author} {\bibfnamefont {C.~J. A.~P.}\
  \bibnamefont {Martins}}, \bibinfo {author} {\bibfnamefont {M.}~\bibnamefont
  {De~Petris}}, \ and\ \bibinfo {author} {\bibfnamefont {L.}~\bibnamefont
  {Lamagna}},\ }\href {\doibase 10.1088/1475-7516/2015/09/011} {\bibfield
  {journal} {\bibinfo  {journal} {JCAP}\ }\textbf {\bibinfo {volume} {1509}},\
  \bibinfo {pages} {011} (\bibinfo {year} {2015})},\ \Eprint
  {http://arxiv.org/abs/1502.07858} {arXiv:1502.07858 [astro-ph.CO]}
  \BibitemShut {NoStop}%
\bibitem [{\citenamefont {{Hurier}}\ \emph {et~al.}(2014)\citenamefont
  {{Hurier}}, \citenamefont {{Aghanim}}, \citenamefont {{Douspis}},\ and\
  \citenamefont {{Pointecouteau}}}]{hurier14}%
  \BibitemOpen
  \bibfield  {author} {\bibinfo {author} {\bibfnamefont {G.}~\bibnamefont
  {{Hurier}}}, \bibinfo {author} {\bibfnamefont {N.}~\bibnamefont {{Aghanim}}},
  \bibinfo {author} {\bibfnamefont {M.}~\bibnamefont {{Douspis}}}, \ and\
  \bibinfo {author} {\bibfnamefont {E.}~\bibnamefont {{Pointecouteau}}},\
  }\href {\doibase 10.1051/0004-6361/201322632} {\bibfield  {journal} {\bibinfo
   {journal} {A. \& A.}\ }\textbf {\bibinfo {volume} {561}},\ \bibinfo {eid}
  {A143} (\bibinfo {year} {2014})},\ \Eprint {http://arxiv.org/abs/1311.4694}
  {arXiv:1311.4694} \BibitemShut {NoStop}%
\bibitem [{\citenamefont {Saro}\ \emph {et~al.}(2014)\citenamefont {Saro} \emph
  {et~al.}}]{saro14}%
  \BibitemOpen
  \bibfield  {author} {\bibinfo {author} {\bibfnamefont {A.}~\bibnamefont
  {Saro}} \emph {et~al.} (\bibinfo {collaboration} {SPT}),\ }\href {\doibase
  10.1093/mnras/stu575} {\bibfield  {journal} {\bibinfo  {journal} {Mon. Not.
  Roy. Astron. Soc.}\ }\textbf {\bibinfo {volume} {440}},\ \bibinfo {pages}
  {2610} (\bibinfo {year} {2014})},\ \Eprint {http://arxiv.org/abs/1312.2462}
  {arXiv:1312.2462 [astro-ph.CO]} \BibitemShut {NoStop}%
\bibitem [{\citenamefont {{Muller}}\ \emph {et~al.}(2013)\citenamefont
  {{Muller}}, \citenamefont {{Beelen}}, \citenamefont {{Black}}, \citenamefont
  {{Curran}}, \citenamefont {{Horellou}}, \citenamefont {{Aalto}},
  \citenamefont {{Combes}}, \citenamefont {{Gu{\'e}lin}},\ and\ \citenamefont
  {{Henkel}}}]{muller13}%
  \BibitemOpen
  \bibfield  {author} {\bibinfo {author} {\bibfnamefont {S.}~\bibnamefont
  {{Muller}}}, \bibinfo {author} {\bibfnamefont {A.}~\bibnamefont {{Beelen}}},
  \bibinfo {author} {\bibfnamefont {J.~H.}\ \bibnamefont {{Black}}}, \bibinfo
  {author} {\bibfnamefont {S.~J.}\ \bibnamefont {{Curran}}}, \bibinfo {author}
  {\bibfnamefont {C.}~\bibnamefont {{Horellou}}}, \bibinfo {author}
  {\bibfnamefont {S.}~\bibnamefont {{Aalto}}}, \bibinfo {author} {\bibfnamefont
  {F.}~\bibnamefont {{Combes}}}, \bibinfo {author} {\bibfnamefont
  {M.}~\bibnamefont {{Gu{\'e}lin}}}, \ and\ \bibinfo {author} {\bibfnamefont
  {C.}~\bibnamefont {{Henkel}}},\ }\href {\doibase 10.1051/0004-6361/201220613}
  {\bibfield  {journal} {\bibinfo  {journal} {A. \& A.}\ }\textbf {\bibinfo
  {volume} {551}},\ \bibinfo {eid} {A109} (\bibinfo {year} {2013})},\ \Eprint
  {http://arxiv.org/abs/1212.5456} {arXiv:1212.5456} \BibitemShut {NoStop}%
\bibitem [{\citenamefont {{Noterdaeme}}\ \emph {et~al.}(2011)\citenamefont
  {{Noterdaeme}}, \citenamefont {{Petitjean}}, \citenamefont {{Srianand}},
  \citenamefont {{Ledoux}},\ and\ \citenamefont {{L{\'o}pez}}}]{noterdaeme11}%
  \BibitemOpen
  \bibfield  {author} {\bibinfo {author} {\bibfnamefont {P.}~\bibnamefont
  {{Noterdaeme}}}, \bibinfo {author} {\bibfnamefont {P.}~\bibnamefont
  {{Petitjean}}}, \bibinfo {author} {\bibfnamefont {R.}~\bibnamefont
  {{Srianand}}}, \bibinfo {author} {\bibfnamefont {C.}~\bibnamefont
  {{Ledoux}}}, \ and\ \bibinfo {author} {\bibfnamefont {S.}~\bibnamefont
  {{L{\'o}pez}}},\ }\href {\doibase 10.1051/0004-6361/201016140} {\bibfield
  {journal} {\bibinfo  {journal} {A. \& A.}\ }\textbf {\bibinfo {volume}
  {526}},\ \bibinfo {eid} {L7} (\bibinfo {year} {2011})},\ \Eprint
  {http://arxiv.org/abs/1012.3164} {arXiv:1012.3164} \BibitemShut {NoStop}%
\bibitem [{\citenamefont {{Cui}}\ \emph {et~al.}(2005)\citenamefont {{Cui}},
  \citenamefont {{Bechtold}}, \citenamefont {{Ge}},\ and\ \citenamefont
  {{Meyer}}}]{cui05}%
  \BibitemOpen
  \bibfield  {author} {\bibinfo {author} {\bibfnamefont {J.}~\bibnamefont
  {{Cui}}}, \bibinfo {author} {\bibfnamefont {J.}~\bibnamefont {{Bechtold}}},
  \bibinfo {author} {\bibfnamefont {J.}~\bibnamefont {{Ge}}}, \ and\ \bibinfo
  {author} {\bibfnamefont {D.~M.}\ \bibnamefont {{Meyer}}},\ }\href {\doibase
  10.1086/444368} {\bibfield  {journal} {\bibinfo  {journal} {Ap. J.}\ }\textbf
  {\bibinfo {volume} {633}},\ \bibinfo {pages} {649} (\bibinfo {year}
  {2005})},\ \Eprint {http://arxiv.org/abs/astro-ph/0506766} {astro-ph/0506766}
  \BibitemShut {NoStop}%
\bibitem [{\citenamefont {{Ge}}\ \emph {et~al.}(2001)\citenamefont {{Ge}},
  \citenamefont {{Bechtold}},\ and\ \citenamefont {{Kulkarni}}}]{ge01}%
  \BibitemOpen
  \bibfield  {author} {\bibinfo {author} {\bibfnamefont {J.}~\bibnamefont
  {{Ge}}}, \bibinfo {author} {\bibfnamefont {J.}~\bibnamefont {{Bechtold}}}, \
  and\ \bibinfo {author} {\bibfnamefont {V.~P.}\ \bibnamefont {{Kulkarni}}},\
  }\href {\doibase 10.1086/318890} {\bibfield  {journal} {\bibinfo  {journal}
  {Ap. J. Lett.}\ }\textbf {\bibinfo {volume} {547}},\ \bibinfo {pages} {L1}
  (\bibinfo {year} {2001})}\BibitemShut {NoStop}%
\bibitem [{\citenamefont {{Srianand}}\ \emph {et~al.}(2000)\citenamefont
  {{Srianand}}, \citenamefont {{Petitjean}},\ and\ \citenamefont
  {{Ledoux}}}]{srianand00}%
  \BibitemOpen
  \bibfield  {author} {\bibinfo {author} {\bibfnamefont {R.}~\bibnamefont
  {{Srianand}}}, \bibinfo {author} {\bibfnamefont {P.}~\bibnamefont
  {{Petitjean}}}, \ and\ \bibinfo {author} {\bibfnamefont {C.}~\bibnamefont
  {{Ledoux}}},\ }\href@noop {} {\bibfield  {journal} {\bibinfo  {journal}
  {Nature}\ }\textbf {\bibinfo {volume} {408}},\ \bibinfo {pages} {931}
  (\bibinfo {year} {2000})},\ \Eprint {http://arxiv.org/abs/astro-ph/0012222}
  {astro-ph/0012222} \BibitemShut {NoStop}%
\bibitem [{\citenamefont {{Srianand}}\ \emph {et~al.}(2008)\citenamefont
  {{Srianand}}, \citenamefont {{Noterdaeme}}, \citenamefont {{Ledoux}},\ and\
  \citenamefont {{Petitjean}}}]{srianand08}%
  \BibitemOpen
  \bibfield  {author} {\bibinfo {author} {\bibfnamefont {R.}~\bibnamefont
  {{Srianand}}}, \bibinfo {author} {\bibfnamefont {P.}~\bibnamefont
  {{Noterdaeme}}}, \bibinfo {author} {\bibfnamefont {C.}~\bibnamefont
  {{Ledoux}}}, \ and\ \bibinfo {author} {\bibfnamefont {P.}~\bibnamefont
  {{Petitjean}}},\ }\href {\doibase 10.1051/0004-6361:200809727} {\bibfield
  {journal} {\bibinfo  {journal} {A. \& A.}\ }\textbf {\bibinfo {volume}
  {482}},\ \bibinfo {pages} {L39} (\bibinfo {year} {2008})}\BibitemShut
  {NoStop}%
\bibitem [{\citenamefont {{Noterdaeme}}\ \emph {et~al.}(2010)\citenamefont
  {{Noterdaeme}}, \citenamefont {{Petitjean}}, \citenamefont {{Ledoux}},
  \citenamefont {{L{\'o}pez}}, \citenamefont {{Srianand}},\ and\ \citenamefont
  {{Vergani}}}]{noterdaeme10}%
  \BibitemOpen
  \bibfield  {author} {\bibinfo {author} {\bibfnamefont {P.}~\bibnamefont
  {{Noterdaeme}}}, \bibinfo {author} {\bibfnamefont {P.}~\bibnamefont
  {{Petitjean}}}, \bibinfo {author} {\bibfnamefont {C.}~\bibnamefont
  {{Ledoux}}}, \bibinfo {author} {\bibfnamefont {S.}~\bibnamefont
  {{L{\'o}pez}}}, \bibinfo {author} {\bibfnamefont {R.}~\bibnamefont
  {{Srianand}}}, \ and\ \bibinfo {author} {\bibfnamefont {S.~D.}\ \bibnamefont
  {{Vergani}}},\ }\href {\doibase 10.1051/0004-6361/201015147} {\bibfield
  {journal} {\bibinfo  {journal} {A. \& A.}\ }\textbf {\bibinfo {volume}
  {523}},\ \bibinfo {eid} {A80} (\bibinfo {year} {2010})},\ \Eprint
  {http://arxiv.org/abs/1008.0637} {arXiv:1008.0637} \BibitemShut {NoStop}%
\bibitem [{\citenamefont {{Molaro}}\ \emph {et~al.}(2002)\citenamefont
  {{Molaro}}, \citenamefont {{Levshakov}}, \citenamefont
  {{Dessauges-Zavadsky}},\ and\ \citenamefont {{D'Odorico}}}]{molaro02}%
  \BibitemOpen
  \bibfield  {author} {\bibinfo {author} {\bibfnamefont {P.}~\bibnamefont
  {{Molaro}}}, \bibinfo {author} {\bibfnamefont {S.~A.}\ \bibnamefont
  {{Levshakov}}}, \bibinfo {author} {\bibfnamefont {M.}~\bibnamefont
  {{Dessauges-Zavadsky}}}, \ and\ \bibinfo {author} {\bibfnamefont
  {S.}~\bibnamefont {{D'Odorico}}},\ }\href {\doibase
  10.1051/0004-6361:20011698} {\bibfield  {journal} {\bibinfo  {journal} {A. \&
  A.}\ }\textbf {\bibinfo {volume} {381}},\ \bibinfo {pages} {L64} (\bibinfo
  {year} {2002})},\ \Eprint {http://arxiv.org/abs/astro-ph/0111589}
  {astro-ph/0111589} \BibitemShut {NoStop}%
\bibitem [{\citenamefont {{Bahcall}}\ and\ \citenamefont
  {{Wolf}}(1968)}]{bahcall68}%
  \BibitemOpen
  \bibfield  {author} {\bibinfo {author} {\bibfnamefont {J.~N.}\ \bibnamefont
  {{Bahcall}}}\ and\ \bibinfo {author} {\bibfnamefont {R.~A.}\ \bibnamefont
  {{Wolf}}},\ }\href {\doibase 10.1086/149589} {\bibfield  {journal} {\bibinfo
  {journal} {Ap. J.}\ }\textbf {\bibinfo {volume} {152}},\ \bibinfo {pages}
  {701} (\bibinfo {year} {1968})}\BibitemShut {NoStop}%
\bibitem [{\citenamefont {{Planck Collaboration}}\ \emph
  {et~al.}(2015{\natexlab{b}})\citenamefont {{Planck Collaboration}},
  \citenamefont {{Ade}}, \citenamefont {{Aghanim}}, \citenamefont {{Arnaud}},
  \citenamefont {{Ashdown}}, \citenamefont {{Aumont}}, \citenamefont
  {{Baccigalupi}}, \citenamefont {{Banday}}, \citenamefont {{Barreiro}},
  \citenamefont {{Barrena}},\ and\ \citenamefont {et~al.}}]{cpp2015-27}%
  \BibitemOpen
  \bibfield  {author} {\bibinfo {author} {\bibnamefont {{Planck
  Collaboration}}}, \bibinfo {author} {\bibfnamefont {P.~A.~R.}\ \bibnamefont
  {{Ade}}}, \bibinfo {author} {\bibfnamefont {N.}~\bibnamefont {{Aghanim}}},
  \bibinfo {author} {\bibfnamefont {M.}~\bibnamefont {{Arnaud}}}, \bibinfo
  {author} {\bibfnamefont {M.}~\bibnamefont {{Ashdown}}}, \bibinfo {author}
  {\bibfnamefont {J.}~\bibnamefont {{Aumont}}}, \bibinfo {author}
  {\bibfnamefont {C.}~\bibnamefont {{Baccigalupi}}}, \bibinfo {author}
  {\bibfnamefont {A.~J.}\ \bibnamefont {{Banday}}}, \bibinfo {author}
  {\bibfnamefont {R.~B.}\ \bibnamefont {{Barreiro}}}, \bibinfo {author}
  {\bibfnamefont {R.}~\bibnamefont {{Barrena}}}, \ and\ \bibinfo {author}
  {\bibnamefont {et~al.}},\ }\href@noop {} {\bibfield  {journal} {\bibinfo
  {journal} {ArXiv e-prints}\ } (\bibinfo {year} {2015}{\natexlab{b}})},\
  \Eprint {http://arxiv.org/abs/1502.01598} {arXiv:1502.01598} \BibitemShut
  {NoStop}%
\bibitem [{\citenamefont {{D'Agostini}}(1999)}]{Dagos1999}%
  \BibitemOpen
  \bibfield  {author} {\bibinfo {author} {\bibfnamefont {G.}~\bibnamefont
  {{D'Agostini}}},\ }\href@noop {} {\bibfield  {journal} {\bibinfo  {journal}
  {ArXiv High Energy Physics - Experiment e-prints}\ } (\bibinfo {year}
  {1999})},\ \Eprint {http://arxiv.org/abs/hep-ex/9910036} {hep-ex/9910036}
  \BibitemShut {NoStop}%
\bibitem [{\citenamefont {Bassett}\ and\ \citenamefont
  {Kunz}(2004)}]{BassettKunz}%
  \BibitemOpen
  \bibfield  {author} {\bibinfo {author} {\bibfnamefont {B.~A.}\ \bibnamefont
  {Bassett}}\ and\ \bibinfo {author} {\bibfnamefont {M.}~\bibnamefont {Kunz}},\
  }\href {\doibase 10.1103/PhysRevD.69.101305} {\bibfield  {journal} {\bibinfo
  {journal} {Phys.Rev.}\ }\textbf {\bibinfo {volume} {D69}},\ \bibinfo {pages}
  {101305} (\bibinfo {year} {2004})},\ \Eprint
  {http://arxiv.org/abs/astro-ph/0312443} {arXiv:astro-ph/0312443 [astro-ph]}
  \BibitemShut {NoStop}%
\bibitem [{\citenamefont {Csaki}\ \emph
  {et~al.}(2002{\natexlab{a}})\citenamefont {Csaki}, \citenamefont {Kaloper},\
  and\ \citenamefont {Terning}}]{Csaki_PRL}%
  \BibitemOpen
  \bibfield  {author} {\bibinfo {author} {\bibfnamefont {C.}~\bibnamefont
  {Csaki}}, \bibinfo {author} {\bibfnamefont {N.}~\bibnamefont {Kaloper}}, \
  and\ \bibinfo {author} {\bibfnamefont {J.}~\bibnamefont {Terning}},\ }\href
  {\doibase 10.1103/PhysRevLett.88.161302} {\bibfield  {journal} {\bibinfo
  {journal} {Phys.Rev.Lett.}\ }\textbf {\bibinfo {volume} {88}},\ \bibinfo
  {pages} {161302} (\bibinfo {year} {2002}{\natexlab{a}})},\ \Eprint
  {http://arxiv.org/abs/hep-ph/0111311} {arXiv:hep-ph/0111311 [hep-ph]}
  \BibitemShut {NoStop}%
\bibitem [{\citenamefont {Csaki}\ \emph
  {et~al.}(2002{\natexlab{b}})\citenamefont {Csaki}, \citenamefont {Kaloper},\
  and\ \citenamefont {Terning}}]{Csaki_PLB}%
  \BibitemOpen
  \bibfield  {author} {\bibinfo {author} {\bibfnamefont {C.}~\bibnamefont
  {Csaki}}, \bibinfo {author} {\bibfnamefont {N.}~\bibnamefont {Kaloper}}, \
  and\ \bibinfo {author} {\bibfnamefont {J.}~\bibnamefont {Terning}},\ }\href
  {\doibase 10.1016/S0370-2693(02)01765-3} {\bibfield  {journal} {\bibinfo
  {journal} {Phys.Lett.}\ }\textbf {\bibinfo {volume} {B535}},\ \bibinfo
  {pages} {33} (\bibinfo {year} {2002}{\natexlab{b}})},\ \Eprint
  {http://arxiv.org/abs/hep-ph/0112212} {arXiv:hep-ph/0112212 [hep-ph]}
  \BibitemShut {NoStop}%
\bibitem [{\citenamefont {Avgoustidis}\ \emph {et~al.}(2010)\citenamefont
  {Avgoustidis}, \citenamefont {Burrage}, \citenamefont {Redondo},
  \citenamefont {Verde},\ and\ \citenamefont {Jimenez}}]{ABRVJ}%
  \BibitemOpen
  \bibfield  {author} {\bibinfo {author} {\bibfnamefont {A.}~\bibnamefont
  {Avgoustidis}}, \bibinfo {author} {\bibfnamefont {C.}~\bibnamefont
  {Burrage}}, \bibinfo {author} {\bibfnamefont {J.}~\bibnamefont {Redondo}},
  \bibinfo {author} {\bibfnamefont {L.}~\bibnamefont {Verde}}, \ and\ \bibinfo
  {author} {\bibfnamefont {R.}~\bibnamefont {Jimenez}},\ }\href {\doibase
  10.1088/1475-7516/2010/10/024} {\bibfield  {journal} {\bibinfo  {journal}
  {JCAP}\ }\textbf {\bibinfo {volume} {1010}},\ \bibinfo {pages} {024}
  (\bibinfo {year} {2010})},\ \Eprint {http://arxiv.org/abs/1004.2053}
  {arXiv:1004.2053 [astro-ph.CO]} \BibitemShut {NoStop}%
\bibitem [{\citenamefont {Burrage}(2008)}]{Bur_Cham}%
  \BibitemOpen
  \bibfield  {author} {\bibinfo {author} {\bibfnamefont {C.}~\bibnamefont
  {Burrage}},\ }\href {\doibase 10.1103/PhysRevD.77.043009} {\bibfield
  {journal} {\bibinfo  {journal} {Phys.Rev.}\ }\textbf {\bibinfo {volume}
  {D77}},\ \bibinfo {pages} {043009} (\bibinfo {year} {2008})},\ \Eprint
  {http://arxiv.org/abs/0711.2966} {arXiv:0711.2966 [astro-ph]} \BibitemShut
  {NoStop}%
\bibitem [{\citenamefont {Avgoustidis}\ \emph {et~al.}(2014)\citenamefont
  {Avgoustidis}, \citenamefont {Martins}, \citenamefont {Monteiro},
  \citenamefont {Vielzeuf},\ and\ \citenamefont {Luzzi}}]{AMMVL}%
  \BibitemOpen
  \bibfield  {author} {\bibinfo {author} {\bibfnamefont {A.}~\bibnamefont
  {Avgoustidis}}, \bibinfo {author} {\bibfnamefont {C.~J. A.~P.}\ \bibnamefont
  {Martins}}, \bibinfo {author} {\bibfnamefont {A.}~\bibnamefont {Monteiro}},
  \bibinfo {author} {\bibfnamefont {P.}~\bibnamefont {Vielzeuf}}, \ and\
  \bibinfo {author} {\bibfnamefont {G.}~\bibnamefont {Luzzi}},\ }\href
  {\doibase 10.1088/1475-7516/2014/06/062} {\bibfield  {journal} {\bibinfo
  {journal} {JCAP}\ }\textbf {\bibinfo {volume} {1406}},\ \bibinfo {pages}
  {062} (\bibinfo {year} {2014})},\ \Eprint {http://arxiv.org/abs/1305.7031}
  {arXiv:1305.7031 [astro-ph.CO]} \BibitemShut {NoStop}%
\bibitem [{\citenamefont {Aguirre}(1999)}]{Aguirre}%
  \BibitemOpen
  \bibfield  {author} {\bibinfo {author} {\bibfnamefont {A.~N.}\ \bibnamefont
  {Aguirre}},\ }\href {\doibase 10.1086/311862} {\bibfield  {journal} {\bibinfo
   {journal} {Astrophys.J.}\ }\textbf {\bibinfo {volume} {512}},\ \bibinfo
  {pages} {L19} (\bibinfo {year} {1999})},\ \Eprint
  {http://arxiv.org/abs/astro-ph/9811316} {arXiv:astro-ph/9811316 [astro-ph]}
  \BibitemShut {NoStop}%
\bibitem [{\citenamefont {Menard}\ \emph {et~al.}(2010)\citenamefont {Menard},
  \citenamefont {Kilbinger},\ and\ \citenamefont {Scranton}}]{Menard}%
  \BibitemOpen
  \bibfield  {author} {\bibinfo {author} {\bibfnamefont {B.}~\bibnamefont
  {Menard}}, \bibinfo {author} {\bibfnamefont {M.}~\bibnamefont {Kilbinger}}, \
  and\ \bibinfo {author} {\bibfnamefont {R.}~\bibnamefont {Scranton}},\
  }\href@noop {} {\bibfield  {journal} {\bibinfo  {journal}
  {Mon.Not.Roy.Astron.Soc.}\ }\textbf {\bibinfo {volume} {406}},\ \bibinfo
  {pages} {1815} (\bibinfo {year} {2010})},\ \Eprint
  {http://arxiv.org/abs/0903.4199} {arXiv:0903.4199 [astro-ph.CO]} \BibitemShut
  {NoStop}%
\bibitem [{\citenamefont {Etherington}(1933)}]{Etherington}%
  \BibitemOpen
  \bibfield  {author} {\bibinfo {author} {\bibfnamefont {J.~M.~H.}\
  \bibnamefont {Etherington}},\ }\href@noop {} {\bibfield  {journal} {\bibinfo
  {journal} {Phil. Mag.}\ }\textbf {\bibinfo {volume} {{\bf 15}}},\ \bibinfo
  {pages} {761} (\bibinfo {year} {1933})}\BibitemShut {NoStop}%
\bibitem [{\citenamefont {Mirizzi}\ \emph {et~al.}(2005)\citenamefont
  {Mirizzi}, \citenamefont {Raffelt},\ and\ \citenamefont
  {Serpico}}]{MirRafSerp_CMB}%
  \BibitemOpen
  \bibfield  {author} {\bibinfo {author} {\bibfnamefont {A.}~\bibnamefont
  {Mirizzi}}, \bibinfo {author} {\bibfnamefont {G.~G.}\ \bibnamefont
  {Raffelt}}, \ and\ \bibinfo {author} {\bibfnamefont {P.~D.}\ \bibnamefont
  {Serpico}},\ }\href {\doibase 10.1103/PhysRevD.72.023501} {\bibfield
  {journal} {\bibinfo  {journal} {Phys.Rev.}\ }\textbf {\bibinfo {volume}
  {D72}},\ \bibinfo {pages} {023501} (\bibinfo {year} {2005})},\ \Eprint
  {http://arxiv.org/abs/astro-ph/0506078} {arXiv:astro-ph/0506078 [astro-ph]}
  \BibitemShut {NoStop}%
\bibitem [{\citenamefont {Ellis}\ \emph {et~al.}(2013)\citenamefont {Ellis},
  \citenamefont {Poltis}, \citenamefont {Uzan},\ and\ \citenamefont
  {Weltman}}]{Ellis2013}%
  \BibitemOpen
  \bibfield  {author} {\bibinfo {author} {\bibfnamefont {G.~F.~R.}\
  \bibnamefont {Ellis}}, \bibinfo {author} {\bibfnamefont {R.}~\bibnamefont
  {Poltis}}, \bibinfo {author} {\bibfnamefont {J.-P.}\ \bibnamefont {Uzan}}, \
  and\ \bibinfo {author} {\bibfnamefont {A.}~\bibnamefont {Weltman}},\ }\href
  {\doibase 10.1103/PhysRevD.87.103530} {\bibfield  {journal} {\bibinfo
  {journal} {Phys. Rev.}\ }\textbf {\bibinfo {volume} {D87}},\ \bibinfo {pages}
  {103530} (\bibinfo {year} {2013})},\ \Eprint {http://arxiv.org/abs/1301.1312}
  {arXiv:1301.1312 [astro-ph.CO]} \BibitemShut {NoStop}%
\bibitem [{\citenamefont {Brax}\ \emph {et~al.}(2013)\citenamefont {Brax},
  \citenamefont {Burrage}, \citenamefont {Davis},\ and\ \citenamefont
  {Gubitosi}}]{Brax2013}%
  \BibitemOpen
  \bibfield  {author} {\bibinfo {author} {\bibfnamefont {P.}~\bibnamefont
  {Brax}}, \bibinfo {author} {\bibfnamefont {C.}~\bibnamefont {Burrage}},
  \bibinfo {author} {\bibfnamefont {A.-C.}\ \bibnamefont {Davis}}, \ and\
  \bibinfo {author} {\bibfnamefont {G.}~\bibnamefont {Gubitosi}},\ }\href
  {\doibase 10.1088/1475-7516/2013/11/001} {\bibfield  {journal} {\bibinfo
  {journal} {JCAP}\ }\textbf {\bibinfo {volume} {1311}},\ \bibinfo {pages}
  {001} (\bibinfo {year} {2013})},\ \Eprint {http://arxiv.org/abs/1306.4168}
  {arXiv:1306.4168 [astro-ph.CO]} \BibitemShut {NoStop}%
\bibitem [{\citenamefont {Dvali}\ and\ \citenamefont
  {Zaldarriaga}(2002)}]{Dvali}%
  \BibitemOpen
  \bibfield  {author} {\bibinfo {author} {\bibfnamefont {G.}~\bibnamefont
  {Dvali}}\ and\ \bibinfo {author} {\bibfnamefont {M.}~\bibnamefont
  {Zaldarriaga}},\ }\href {\doibase 10.1103/PhysRevLett.88.091303} {\bibfield
  {journal} {\bibinfo  {journal} {Phys.Rev.Lett.}\ }\textbf {\bibinfo {volume}
  {88}},\ \bibinfo {pages} {091303} (\bibinfo {year} {2002})},\ \Eprint
  {http://arxiv.org/abs/hep-ph/0108217} {arXiv:hep-ph/0108217 [hep-ph]}
  \BibitemShut {NoStop}%
\bibitem [{\citenamefont {Hees}\ \emph {et~al.}(2014)\citenamefont {Hees},
  \citenamefont {Minazzoli},\ and\ \citenamefont {Larena}}]{Hees14}%
  \BibitemOpen
  \bibfield  {author} {\bibinfo {author} {\bibfnamefont {A.}~\bibnamefont
  {Hees}}, \bibinfo {author} {\bibfnamefont {O.}~\bibnamefont {Minazzoli}}, \
  and\ \bibinfo {author} {\bibfnamefont {J.}~\bibnamefont {Larena}},\ }\href
  {\doibase 10.1103/PhysRevD.90.124064} {\bibfield  {journal} {\bibinfo
  {journal} {Phys.Rev.}\ }\textbf {\bibinfo {volume} {D90}},\ \bibinfo {pages}
  {124064} (\bibinfo {year} {2014})},\ \Eprint {http://arxiv.org/abs/1406.6187}
  {arXiv:1406.6187 [astro-ph.CO]} \BibitemShut {NoStop}%
\bibitem [{\citenamefont {Betoule}\ \emph {et~al.}(2014)\citenamefont {Betoule}
  \emph {et~al.}}]{Betoule2014}%
  \BibitemOpen
  \bibfield  {author} {\bibinfo {author} {\bibfnamefont {M.}~\bibnamefont
  {Betoule}} \emph {et~al.} (\bibinfo {collaboration} {SDSS}),\ }\href
  {\doibase 10.1051/0004-6361/201423413} {\bibfield  {journal} {\bibinfo
  {journal} {Astron.Astrophys.}\ }\textbf {\bibinfo {volume} {568}},\ \bibinfo
  {pages} {A22} (\bibinfo {year} {2014})},\ \Eprint
  {http://arxiv.org/abs/1401.4064} {arXiv:1401.4064 [astro-ph.CO]} \BibitemShut
  {NoStop}%
\bibitem [{\citenamefont {Stern}\ \emph {et~al.}(2010)\citenamefont {Stern},
  \citenamefont {Jimenez}, \citenamefont {Verde}, \citenamefont
  {Kamionkowski},\ and\ \citenamefont {Stanford}}]{Stern2009}%
  \BibitemOpen
  \bibfield  {author} {\bibinfo {author} {\bibfnamefont {D.}~\bibnamefont
  {Stern}}, \bibinfo {author} {\bibfnamefont {R.}~\bibnamefont {Jimenez}},
  \bibinfo {author} {\bibfnamefont {L.}~\bibnamefont {Verde}}, \bibinfo
  {author} {\bibfnamefont {M.}~\bibnamefont {Kamionkowski}}, \ and\ \bibinfo
  {author} {\bibfnamefont {S.~A.}\ \bibnamefont {Stanford}},\ }\href {\doibase
  10.1088/1475-7516/2010/02/008} {\bibfield  {journal} {\bibinfo  {journal}
  {JCAP}\ }\textbf {\bibinfo {volume} {1002}},\ \bibinfo {pages} {008}
  (\bibinfo {year} {2010})},\ \Eprint {http://arxiv.org/abs/0907.3149}
  {arXiv:0907.3149 [astro-ph.CO]} \BibitemShut {NoStop}%
\bibitem [{\citenamefont {Simon}\ \emph {et~al.}(2005)\citenamefont {Simon},
  \citenamefont {Verde},\ and\ \citenamefont {Jimenez}}]{Simon2004}%
  \BibitemOpen
  \bibfield  {author} {\bibinfo {author} {\bibfnamefont {J.}~\bibnamefont
  {Simon}}, \bibinfo {author} {\bibfnamefont {L.}~\bibnamefont {Verde}}, \ and\
  \bibinfo {author} {\bibfnamefont {R.}~\bibnamefont {Jimenez}},\ }\href
  {\doibase 10.1103/PhysRevD.71.123001} {\bibfield  {journal} {\bibinfo
  {journal} {Phys.Rev.}\ }\textbf {\bibinfo {volume} {D71}},\ \bibinfo {pages}
  {123001} (\bibinfo {year} {2005})},\ \Eprint
  {http://arxiv.org/abs/astro-ph/0412269} {arXiv:astro-ph/0412269 [astro-ph]}
  \BibitemShut {NoStop}%
\bibitem [{\citenamefont {Jimenez}\ \emph {et~al.}(2003)\citenamefont
  {Jimenez}, \citenamefont {Verde}, \citenamefont {Treu},\ and\ \citenamefont
  {Stern}}]{Jimenez2003}%
  \BibitemOpen
  \bibfield  {author} {\bibinfo {author} {\bibfnamefont {R.}~\bibnamefont
  {Jimenez}}, \bibinfo {author} {\bibfnamefont {L.}~\bibnamefont {Verde}},
  \bibinfo {author} {\bibfnamefont {T.}~\bibnamefont {Treu}}, \ and\ \bibinfo
  {author} {\bibfnamefont {D.}~\bibnamefont {Stern}},\ }\href {\doibase
  10.1086/376595} {\bibfield  {journal} {\bibinfo  {journal} {Astrophys.J.}\
  }\textbf {\bibinfo {volume} {593}},\ \bibinfo {pages} {622} (\bibinfo {year}
  {2003})},\ \Eprint {http://arxiv.org/abs/astro-ph/0302560}
  {arXiv:astro-ph/0302560 [astro-ph]} \BibitemShut {NoStop}%
\bibitem [{\citenamefont {Moresco}\ \emph {et~al.}(2012)\citenamefont
  {Moresco}, \citenamefont {Cimatti}, \citenamefont {Jimenez}, \citenamefont
  {Pozzetti}, \citenamefont {Zamorani} \emph {et~al.}}]{Moresco2012}%
  \BibitemOpen
  \bibfield  {author} {\bibinfo {author} {\bibfnamefont {M.}~\bibnamefont
  {Moresco}}, \bibinfo {author} {\bibfnamefont {A.}~\bibnamefont {Cimatti}},
  \bibinfo {author} {\bibfnamefont {R.}~\bibnamefont {Jimenez}}, \bibinfo
  {author} {\bibfnamefont {L.}~\bibnamefont {Pozzetti}}, \bibinfo {author}
  {\bibfnamefont {G.}~\bibnamefont {Zamorani}},  \emph {et~al.},\ }\href
  {\doibase 10.1088/1475-7516/2012/08/006} {\bibfield  {journal} {\bibinfo
  {journal} {JCAP}\ }\textbf {\bibinfo {volume} {1208}},\ \bibinfo {pages}
  {006} (\bibinfo {year} {2012})},\ \Eprint {http://arxiv.org/abs/1201.3609}
  {arXiv:1201.3609 [astro-ph.CO]} \BibitemShut {NoStop}%
\bibitem [{\citenamefont {Blake}\ \emph {et~al.}(2012)\citenamefont {Blake},
  \citenamefont {Brough}, \citenamefont {Colless}, \citenamefont {Contreras},
  \citenamefont {Couch} \emph {et~al.}}]{Blake2012}%
  \BibitemOpen
  \bibfield  {author} {\bibinfo {author} {\bibfnamefont {C.}~\bibnamefont
  {Blake}}, \bibinfo {author} {\bibfnamefont {S.}~\bibnamefont {Brough}},
  \bibinfo {author} {\bibfnamefont {M.}~\bibnamefont {Colless}}, \bibinfo
  {author} {\bibfnamefont {C.}~\bibnamefont {Contreras}}, \bibinfo {author}
  {\bibfnamefont {W.}~\bibnamefont {Couch}},  \emph {et~al.},\ }\href {\doibase
  10.1111/j.1365-2966.2012.21473.x} {\bibfield  {journal} {\bibinfo  {journal}
  {Mon.Not.Roy.Astron.Soc.}\ }\textbf {\bibinfo {volume} {425}},\ \bibinfo
  {pages} {405} (\bibinfo {year} {2012})},\ \Eprint
  {http://arxiv.org/abs/1204.3674} {arXiv:1204.3674 [astro-ph.CO]} \BibitemShut
  {NoStop}%
\bibitem [{\citenamefont {Xu}\ \emph {et~al.}(2013)\citenamefont {Xu},
  \citenamefont {Cuesta}, \citenamefont {Padmanabhan}, \citenamefont
  {Eisenstein},\ and\ \citenamefont {McBride}}]{Xu2012}%
  \BibitemOpen
  \bibfield  {author} {\bibinfo {author} {\bibfnamefont {X.}~\bibnamefont
  {Xu}}, \bibinfo {author} {\bibfnamefont {A.~J.}\ \bibnamefont {Cuesta}},
  \bibinfo {author} {\bibfnamefont {N.}~\bibnamefont {Padmanabhan}}, \bibinfo
  {author} {\bibfnamefont {D.~J.}\ \bibnamefont {Eisenstein}}, \ and\ \bibinfo
  {author} {\bibfnamefont {C.~K.}\ \bibnamefont {McBride}},\ }\href {\doibase
  10.1093/mnras/stt379} {\bibfield  {journal} {\bibinfo  {journal} {Mon. Not.
  Roy. Astron. Soc.}\ }\textbf {\bibinfo {volume} {431}},\ \bibinfo {pages}
  {2834} (\bibinfo {year} {2013})},\ \Eprint {http://arxiv.org/abs/1206.6732}
  {arXiv:1206.6732 [astro-ph.CO]} \BibitemShut {NoStop}%
\bibitem [{\citenamefont {Anderson}\ \emph {et~al.}(2014)\citenamefont
  {Anderson} \emph {et~al.}}]{Anderson2013}%
  \BibitemOpen
  \bibfield  {author} {\bibinfo {author} {\bibfnamefont {L.}~\bibnamefont
  {Anderson}} \emph {et~al.} (\bibinfo {collaboration} {BOSS}),\ }\href
  {\doibase 10.1093/mnras/stu523} {\bibfield  {journal} {\bibinfo  {journal}
  {Mon.Not.Roy.Astron.Soc.}\ }\textbf {\bibinfo {volume} {441}},\ \bibinfo
  {pages} {24} (\bibinfo {year} {2014})},\ \Eprint
  {http://arxiv.org/abs/1312.4877} {arXiv:1312.4877 [astro-ph.CO]} \BibitemShut
  {NoStop}%
\bibitem [{\citenamefont {Delubac}\ \emph {et~al.}(2015)\citenamefont {Delubac}
  \emph {et~al.}}]{Delubac2014}%
  \BibitemOpen
  \bibfield  {author} {\bibinfo {author} {\bibfnamefont {T.}~\bibnamefont
  {Delubac}} \emph {et~al.} (\bibinfo {collaboration} {BOSS}),\ }\href
  {\doibase 10.1051/0004-6361/201423969} {\bibfield  {journal} {\bibinfo
  {journal} {Astron.Astrophys.}\ }\textbf {\bibinfo {volume} {574}},\ \bibinfo
  {pages} {A59} (\bibinfo {year} {2015})},\ \Eprint
  {http://arxiv.org/abs/1404.1801} {arXiv:1404.1801 [astro-ph.CO]} \BibitemShut
  {NoStop}%
\bibitem [{\citenamefont {Andr\'e}\ \emph {et~al.}(2014)\citenamefont {Andr\'e}
  \emph {et~al.}}]{PRISM}%
  \BibitemOpen
  \bibfield  {author} {\bibinfo {author} {\bibfnamefont {P.}~\bibnamefont
  {Andr\'e}} \emph {et~al.} (\bibinfo {collaboration} {PRISM}),\ }\href
  {\doibase 10.1088/1475-7516/2014/02/006} {\bibfield  {journal} {\bibinfo
  {journal} {JCAP}\ }\textbf {\bibinfo {volume} {1402}},\ \bibinfo {pages}
  {006} (\bibinfo {year} {2014})},\ \Eprint {http://arxiv.org/abs/1310.1554}
  {arXiv:1310.1554 [astro-ph.CO]} \BibitemShut {NoStop}%
\bibitem [{\citenamefont {Fish}\ \emph {et~al.}(2013)\citenamefont {Fish},
  \citenamefont {Alef}, \citenamefont {Anderson}, \citenamefont {Asada},
  \citenamefont {Baudry} \emph {et~al.}}]{ALMA1}%
  \BibitemOpen
  \bibfield  {author} {\bibinfo {author} {\bibfnamefont {V.}~\bibnamefont
  {Fish}}, \bibinfo {author} {\bibfnamefont {W.}~\bibnamefont {Alef}}, \bibinfo
  {author} {\bibfnamefont {J.}~\bibnamefont {Anderson}}, \bibinfo {author}
  {\bibfnamefont {K.}~\bibnamefont {Asada}}, \bibinfo {author} {\bibfnamefont
  {A.}~\bibnamefont {Baudry}},  \emph {et~al.},\ }\href@noop {} {\  (\bibinfo
  {year} {2013})},\ \Eprint {http://arxiv.org/abs/1309.3519} {arXiv:1309.3519
  [astro-ph.IM]} \BibitemShut {NoStop}%
\bibitem [{\citenamefont {Tilanus}\ \emph {et~al.}(2014)\citenamefont
  {Tilanus}, \citenamefont {Krichbaum}, \citenamefont {Zensus}, \citenamefont
  {Baudry}, \citenamefont {Bremer} \emph {et~al.}}]{ALMA2}%
  \BibitemOpen
  \bibfield  {author} {\bibinfo {author} {\bibfnamefont {R.}~\bibnamefont
  {Tilanus}}, \bibinfo {author} {\bibfnamefont {T.}~\bibnamefont {Krichbaum}},
  \bibinfo {author} {\bibfnamefont {J.}~\bibnamefont {Zensus}}, \bibinfo
  {author} {\bibfnamefont {A.}~\bibnamefont {Baudry}}, \bibinfo {author}
  {\bibfnamefont {M.}~\bibnamefont {Bremer}},  \emph {et~al.},\ }\href@noop {}
  {\  (\bibinfo {year} {2014})},\ \Eprint {http://arxiv.org/abs/1406.4650}
  {arXiv:1406.4650 [astro-ph.IM]} \BibitemShut {NoStop}%
\bibitem [{\citenamefont {{Pepe}}\ \emph {et~al.}(2013)\citenamefont {{Pepe}},
  \citenamefont {{Cristiani}}, \citenamefont {{Rebolo}}, \citenamefont
  {{Santos}}, \citenamefont {{Dekker}}, \citenamefont {{M{\'e}gevand}},
  \citenamefont {{Zerbi}}, \citenamefont {{Cabral}}, \citenamefont {{Molaro}},
  \citenamefont {{Di Marcantonio}}, \citenamefont {{Abreu}}, \citenamefont
  {{Affolter}}, \citenamefont {{Aliverti}}, \citenamefont {{Allende Prieto}},
  \citenamefont {{Amate}}, \citenamefont {{Avila}}, \citenamefont {{Baldini}},
  \citenamefont {{Bristow}}, \citenamefont {{Broeg}}, \citenamefont {{Cirami}},
  \citenamefont {{Coelho}}, \citenamefont {{Conconi}}, \citenamefont
  {{Coretti}}, \citenamefont {{Cupani}}, \citenamefont {{D'Odorico}},
  \citenamefont {{De Caprio}}, \citenamefont {{Delabre}}, \citenamefont
  {{Dorn}}, \citenamefont {{Figueira}}, \citenamefont {{Fragoso}},
  \citenamefont {{Galeotta}}, \citenamefont {{Genolet}}, \citenamefont
  {{Gomes}}, \citenamefont {{Gonz{\'a}lez Hern{\'a}ndez}}, \citenamefont
  {{Hughes}}, \citenamefont {{Iwert}}, \citenamefont {{Kerber}}, \citenamefont
  {{Landoni}}, \citenamefont {{Lizon}}, \citenamefont {{Lovis}}, \citenamefont
  {{Maire}}, \citenamefont {{Mannetta}}, \citenamefont {{Martins}},
  \citenamefont {{Monteiro}}, \citenamefont {{Oliveira}}, \citenamefont
  {{Poretti}}, \citenamefont {{Rasilla}}, \citenamefont {{Riva}}, \citenamefont
  {{Santana Tschudi}}, \citenamefont {{Santos}}, \citenamefont {{Sosnowska}},
  \citenamefont {{Sousa}}, \citenamefont {{Span{\`o}}}, \citenamefont
  {{Tenegi}}, \citenamefont {{Toso}}, \citenamefont {{Vanzella}}, \citenamefont
  {{Viel}},\ and\ \citenamefont {{Zapatero Osorio}}}]{ESPRESSO}%
  \BibitemOpen
  \bibfield  {author} {\bibinfo {author} {\bibfnamefont {F.}~\bibnamefont
  {{Pepe}}}, \bibinfo {author} {\bibfnamefont {S.}~\bibnamefont {{Cristiani}}},
  \bibinfo {author} {\bibfnamefont {R.}~\bibnamefont {{Rebolo}}}, \bibinfo
  {author} {\bibfnamefont {N.~C.}\ \bibnamefont {{Santos}}}, \bibinfo {author}
  {\bibfnamefont {H.}~\bibnamefont {{Dekker}}}, \bibinfo {author}
  {\bibfnamefont {D.}~\bibnamefont {{M{\'e}gevand}}}, \bibinfo {author}
  {\bibfnamefont {F.~M.}\ \bibnamefont {{Zerbi}}}, \bibinfo {author}
  {\bibfnamefont {A.}~\bibnamefont {{Cabral}}}, \bibinfo {author}
  {\bibfnamefont {P.}~\bibnamefont {{Molaro}}}, \bibinfo {author}
  {\bibfnamefont {P.}~\bibnamefont {{Di Marcantonio}}}, \bibinfo {author}
  {\bibfnamefont {M.}~\bibnamefont {{Abreu}}}, \bibinfo {author} {\bibfnamefont
  {M.}~\bibnamefont {{Affolter}}}, \bibinfo {author} {\bibfnamefont
  {M.}~\bibnamefont {{Aliverti}}}, \bibinfo {author} {\bibfnamefont
  {C.}~\bibnamefont {{Allende Prieto}}}, \bibinfo {author} {\bibfnamefont
  {M.}~\bibnamefont {{Amate}}}, \bibinfo {author} {\bibfnamefont
  {G.}~\bibnamefont {{Avila}}}, \bibinfo {author} {\bibfnamefont
  {V.}~\bibnamefont {{Baldini}}}, \bibinfo {author} {\bibfnamefont
  {P.}~\bibnamefont {{Bristow}}}, \bibinfo {author} {\bibfnamefont
  {C.}~\bibnamefont {{Broeg}}}, \bibinfo {author} {\bibfnamefont
  {R.}~\bibnamefont {{Cirami}}}, \bibinfo {author} {\bibfnamefont
  {J.}~\bibnamefont {{Coelho}}}, \bibinfo {author} {\bibfnamefont
  {P.}~\bibnamefont {{Conconi}}}, \bibinfo {author} {\bibfnamefont
  {I.}~\bibnamefont {{Coretti}}}, \bibinfo {author} {\bibfnamefont
  {G.}~\bibnamefont {{Cupani}}}, \bibinfo {author} {\bibfnamefont
  {V.}~\bibnamefont {{D'Odorico}}}, \bibinfo {author} {\bibfnamefont
  {V.}~\bibnamefont {{De Caprio}}}, \bibinfo {author} {\bibfnamefont
  {B.}~\bibnamefont {{Delabre}}}, \bibinfo {author} {\bibfnamefont
  {R.}~\bibnamefont {{Dorn}}}, \bibinfo {author} {\bibfnamefont
  {P.}~\bibnamefont {{Figueira}}}, \bibinfo {author} {\bibfnamefont
  {A.}~\bibnamefont {{Fragoso}}}, \bibinfo {author} {\bibfnamefont
  {S.}~\bibnamefont {{Galeotta}}}, \bibinfo {author} {\bibfnamefont
  {L.}~\bibnamefont {{Genolet}}}, \bibinfo {author} {\bibfnamefont
  {R.}~\bibnamefont {{Gomes}}}, \bibinfo {author} {\bibfnamefont {J.~I.}\
  \bibnamefont {{Gonz{\'a}lez Hern{\'a}ndez}}}, \bibinfo {author}
  {\bibfnamefont {I.}~\bibnamefont {{Hughes}}}, \bibinfo {author}
  {\bibfnamefont {O.}~\bibnamefont {{Iwert}}}, \bibinfo {author} {\bibfnamefont
  {F.}~\bibnamefont {{Kerber}}}, \bibinfo {author} {\bibfnamefont
  {M.}~\bibnamefont {{Landoni}}}, \bibinfo {author} {\bibfnamefont {J.-L.}\
  \bibnamefont {{Lizon}}}, \bibinfo {author} {\bibfnamefont {C.}~\bibnamefont
  {{Lovis}}}, \bibinfo {author} {\bibfnamefont {C.}~\bibnamefont {{Maire}}},
  \bibinfo {author} {\bibfnamefont {M.}~\bibnamefont {{Mannetta}}}, \bibinfo
  {author} {\bibfnamefont {C.}~\bibnamefont {{Martins}}}, \bibinfo {author}
  {\bibfnamefont {M.~A.}\ \bibnamefont {{Monteiro}}}, \bibinfo {author}
  {\bibfnamefont {A.}~\bibnamefont {{Oliveira}}}, \bibinfo {author}
  {\bibfnamefont {E.}~\bibnamefont {{Poretti}}}, \bibinfo {author}
  {\bibfnamefont {J.~L.}\ \bibnamefont {{Rasilla}}}, \bibinfo {author}
  {\bibfnamefont {M.}~\bibnamefont {{Riva}}}, \bibinfo {author} {\bibfnamefont
  {S.}~\bibnamefont {{Santana Tschudi}}}, \bibinfo {author} {\bibfnamefont
  {P.}~\bibnamefont {{Santos}}}, \bibinfo {author} {\bibfnamefont
  {D.}~\bibnamefont {{Sosnowska}}}, \bibinfo {author} {\bibfnamefont
  {S.}~\bibnamefont {{Sousa}}}, \bibinfo {author} {\bibfnamefont
  {P.}~\bibnamefont {{Span{\`o}}}}, \bibinfo {author} {\bibfnamefont
  {F.}~\bibnamefont {{Tenegi}}}, \bibinfo {author} {\bibfnamefont
  {G.}~\bibnamefont {{Toso}}}, \bibinfo {author} {\bibfnamefont
  {E.}~\bibnamefont {{Vanzella}}}, \bibinfo {author} {\bibfnamefont
  {M.}~\bibnamefont {{Viel}}}, \ and\ \bibinfo {author} {\bibfnamefont {M.~R.}\
  \bibnamefont {{Zapatero Osorio}}},\ }\href@noop {} {\bibfield  {journal}
  {\bibinfo  {journal} {The Messenger}\ }\textbf {\bibinfo {volume} {153}},\
  \bibinfo {pages} {6} (\bibinfo {year} {2013})}\BibitemShut {NoStop}%
\bibitem [{\citenamefont {Liske}\ \emph {et~al.}(2014)\citenamefont {Liske}
  \emph {et~al.}}]{EELT}%
  \BibitemOpen
  \bibfield  {author} {\bibinfo {author} {\bibfnamefont {J.}~\bibnamefont
  {Liske}} \emph {et~al.},\ }\href@noop {} {\bibfield  {journal} {\bibinfo
  {journal} {{Top Level Requirements For ELT-HIRES}}\ ,\ \bibinfo {pages}
  {Document ESO 204697 Version 1}} (\bibinfo {year} {2014})}\BibitemShut
  {NoStop}%
\bibitem [{\citenamefont {{Kre{\l}owski}}\ \emph {et~al.}(2012)\citenamefont
  {{Kre{\l}owski}}, \citenamefont {{Galazutdinov}},\ and\ \citenamefont
  {{Gnaci{\'n}ski}}}]{KREL}%
  \BibitemOpen
  \bibfield  {author} {\bibinfo {author} {\bibfnamefont {J.}~\bibnamefont
  {{Kre{\l}owski}}}, \bibinfo {author} {\bibfnamefont {G.}~\bibnamefont
  {{Galazutdinov}}}, \ and\ \bibinfo {author} {\bibfnamefont {P.}~\bibnamefont
  {{Gnaci{\'n}ski}}},\ }\href {\doibase 10.1002/asna.201111708} {\bibfield
  {journal} {\bibinfo  {journal} {Astronomische Nachrichten}\ }\textbf
  {\bibinfo {volume} {333}},\ \bibinfo {pages} {627} (\bibinfo {year}
  {2012})}\BibitemShut {NoStop}%
\bibitem [{\citenamefont {Martins}(2015)}]{GRG}%
  \BibitemOpen
  \bibfield  {author} {\bibinfo {author} {\bibfnamefont {C.~J. A.~P.}\
  \bibnamefont {Martins}},\ }\href {\doibase 10.1007/s10714-014-1843-7}
  {\bibfield  {journal} {\bibinfo  {journal} {Gen.Rel.Grav.}\ }\textbf
  {\bibinfo {volume} {47}},\ \bibinfo {pages} {1843} (\bibinfo {year}
  {2015})},\ \Eprint {http://arxiv.org/abs/1412.0108} {arXiv:1412.0108
  [astro-ph.CO]} \BibitemShut {NoStop}%
\end{thebibliography}%
\end{document}